\documentclass[aps,prl,twocolumn,superscriptaddress,longbibliography]{revtex4-2}%
\usepackage{graphicx}
\usepackage{float}
\usepackage{dcolumn}
\usepackage{bm,color}
\usepackage{amsmath,amssymb,dsfont,amstext,amsfonts}
\usepackage{extarrows}
\usepackage[colorlinks=true,linkcolor=blue,urlcolor=blue,citecolor=blue]{hyperref}
\usepackage{xcolor}
\usepackage{wasysym}
\usepackage{mathtools}
\usepackage{bbold}

\usepackage[normalem]{ulem} %makes underlining possible: \uline \uwave \sout
\usepackage{color}%makes \textcolor{grey}{text} available

 \usepackage[final]{changes}
\makeatletter
\newcommand{\btriangle}{\mathpalette\btriangle@\relax}
\newcommand{\btriangle@}[2]{%
  \begingroup
  \sbox\z@{$\m@th#1\triangle$}%
  \makebox[\wd\z@]{%
      \resizebox{1.1\wd\z@}{0.96\ht\z@}{%
        $\m@th#1\blacktriangle$%
      }%
  }%
  \endgroup 
  }
\makeatother
\makeatletter
\newcommand{\striangle}{\mathpalette\striangle@\relax}
\newcommand{\striangle@}[2]{%
  \begingroup
  \sbox\z@{$\m@th#1\triangle$}%
  \makebox[\wd\z@]{%
      \resizebox{0.8\wd\z@}{0.85\ht\z@}{%
        $\m@th#1\triangle$%
      }%
  }%
  \endgroup 
  }
\makeatother

\newcommand{\bk}{\bm{k}}

\newcommand{\parag}[1]{\vspace{0.3em}{\it #1.}}

\def\ket#1{\mathinner{|{#1}\rangle}}

\def\braket#1{\mathinner{\langle{#1}\rangle}}

\begin{document}
\title{Interferometry of non-Abelian band singularities and Euler class topology}
\author{Oliver Breach}
\affiliation{TCM Group, Cavendish Laboratory, University of Cambridge, JJ Thomson Avenue, Cambridge CB3 0HE, United Kingdom\looseness=-1}
\affiliation{Rudolf Peierls Centre for Theoretical Physics, Parks Road, Oxford, OX1 3PU, United Kingdom\looseness=-1}
\author{Robert-Jan Slager}
\affiliation{TCM Group, Cavendish Laboratory, University of Cambridge, JJ Thomson Avenue, Cambridge CB3 0HE, United Kingdom\looseness=-1}
\author{F.~Nur \"{U}nal}
\email{fnu20@cam.ac.uk}
\affiliation{TCM Group, Cavendish Laboratory, University of Cambridge, JJ Thomson Avenue, Cambridge CB3 0HE, United Kingdom\looseness=-1}

%%TC:ignore
%\date{\today}
\begin{abstract}
In systems with a real Bloch Hamiltonian band nodes can be characterised by a non-Abelian frame-rotation charge. The ability of these band nodes to annihilate pairwise is path dependent, since by braiding nodes in adjacent gaps the sign of their charges can be changed. 
Here, we theoretically construct and numerically confirm two concrete methods to experimentally probe these non-Abelian braiding processes and charges in ultracold atomic systems. 
We consider a coherent superposition of two bands that can be created by moving atoms through the band singularities at some angle in momentum space. Analyzing the dependency of excitations on the frame charges, we demonstrate an interferometry scheme passing through two band nodes, which reveals the relative frame charges and allows for measuring the multi-gap topological invariant. The second method relies on a single wavepacket probing two nodes sequentially, where the frame charges can be determined from the band populations. Our results present a feasible avenue for measuring non-Abelian charges of band nodes and the direct experimental verification of braiding procedures, which can be applied in a variety of settings including the recently discovered anomalous non-Abelian phases arising under periodic driving.
\end{abstract}
\maketitle
%%TC:endignore

%%%%%%%%%%%%%%%%%%%%%%%%%%%%%%%%%%%%%%%%%%%%%%%%%%%%%%%%%
\parag{Introduction}---
Within the active field of topological insulators and semimetals~\cite{Rmp1,Rmp2,Weylrmp}, the role of on-site ten-fold way symmetries, such as time reversal, particle-hole and chiral, as well as crystalline symmetries are by now rather uniformly understood in and out of equilibrium~\cite{clas1,Clas2, SchnyderClass, Clas3,Clas4, Clas5,Kitaev, bouhon2018wilson, Ft1, Roy17_PRB, Rudner13_PRX,Kitagawa10_PRB,LieuCooper20_PRL}. Recently there has been new progress in terms of multi-gap topological phases~\cite{ Bouhon2020nonabelian,Ahn2019nielsen, bouhon2020geometric, Jiang2021acoustic, Guo1Dexp}, %. In such systems, 
where band subspaces (sets of isolated bands) can attain non-trivial invariants that do not a priori depend on the symmetry eigenvalues of the bands at high  symmetry points and, hence, fall outside of all hitherto-known classifications~\cite{Clas3,Clas4, Clas5, Kitaev}. When a system can be represented in terms of a real-valued Hamiltonian due to the presence of $\mathcal{C}_2 \mathcal{T}$ or $\mathcal{P} \mathcal{T}$ (two-fold rotations or parity and time-reversal) symmetry, band degeneracies can carry non-Abelian frame charges~\cite{Wu1273}. The sign of these frame charges can be changed upon braiding band nodes in the momentum space, where the obstruction to annihilate the similarly-valued charges is quantified by a multi-gap invariant, the Euler class~\cite{Ahn2019nielsen, Bouhon2020nonabelian, bouhon2020geometric}.
%Interestingly, braiding nodes residing in adjacent band gaps can ensure that a band subspace features similarly-valued, rather than opposite, nodal charges, whose obstruction to annihilate is quantified by a multi-gap invariant, the Euler class~\cite{}. 
Multi-gap considerations away from equilibrium have revealed even more exotic phenomena, such as novel quench signatures~\cite{Unal2020_PRL,zhao2022observation}, optical responses~\cite{jankowski2023optical,Avdoshkin2023,kwon2023quantum, bouhon2023quantum} and anomalous non-Abelian phases that can exclusively arise under periodic driving~\cite{Slager2024}.
%that have been swiftly observed

Band nodes also play an important role in single-gap topological phases. Weyl nodes act as sources of Berry flux~\cite{Vanderbilt2018}, while higher order degeneracies require stabilization of crystalline symmetries and are often associated with topological invariants. 
Individually, their properties can be probed in different experiments~\cite{Cooper19_RMP,Duca2015,Tarnowski2017,Brown2022,Tarruell2012,MahmoodGedik16_NatPhys}. For example, the $\pi$-Berry flux of a Dirac cone has been measured through atomic momentum-space interferometry in optical lattices~\cite{Duca2015}, and the winding of isolated linear or quadratic band touchings has been observed via exciting atoms by moving them through nodes~\cite{Brown2022}.
%Advances in ultracold atoms have allowed for directly probing these topological nodes individually, such as with interferometry to detect the Aharonov-Bohm flux of a single Dirac cone~\cite{} and measuring the topological winding around singly or doubly charged nodes via excitations~\cite{Brown2022}. 
Upon being elevated to multi-gap topologies, the hallmark gained by band singularities is non-Abelian frame charges that can induce a nontrivial Euler class. Consequently, a fundamental question arises whether there exists observable signatures of these non-Abelian charges, their braiding and this invariant. %taking values in the quaternion group. %The frame charges nonetheless correspond to $\pm\pi$ phases within a gap and, hence, cannot be captured by conventional interferometry techniques which remain insensitive to the sign change.

We here address this question, providing an essential link to experiments. Whilst frame charges correspond to $\pm\pi$ vortices {\it within} a gap, distinguishing them is difficult because it amounts to discriminating between zero and $2\pi$ phase windings of the frame after braiding. 
We demonstrate that these two cases impose certain constructive or destructive interference of atoms passing through different Euler nodes. Secondly, we construct another protocol for sequential excitation of atoms by moving them through two band nodes consecutively where the frame charges can be detected in the phase shift of resulting oscillations in band populations.
%We overcome this challenge and construct two different protocols to access this relative sign. In particular, the first involving two paths which each pass through a node, and secondly, a sequential interferometry of excitations resulting from atoms entering, turning and exiting two band nodes in the Euler class consecutively. 
Our results reveal a physical manifestation of non-Abelian charges and their relative sign changes upon braiding in momentum space, imposing a feasible route to directly observe signatures of multi-gap band topology in experiments.
%the physical signatures of the non-Abelian charges, but also pave the way for directly observing these frame charges and consequences of the momentum-space braiding.

\begin{figure}
\centering
\includegraphics[width=8.6cm]{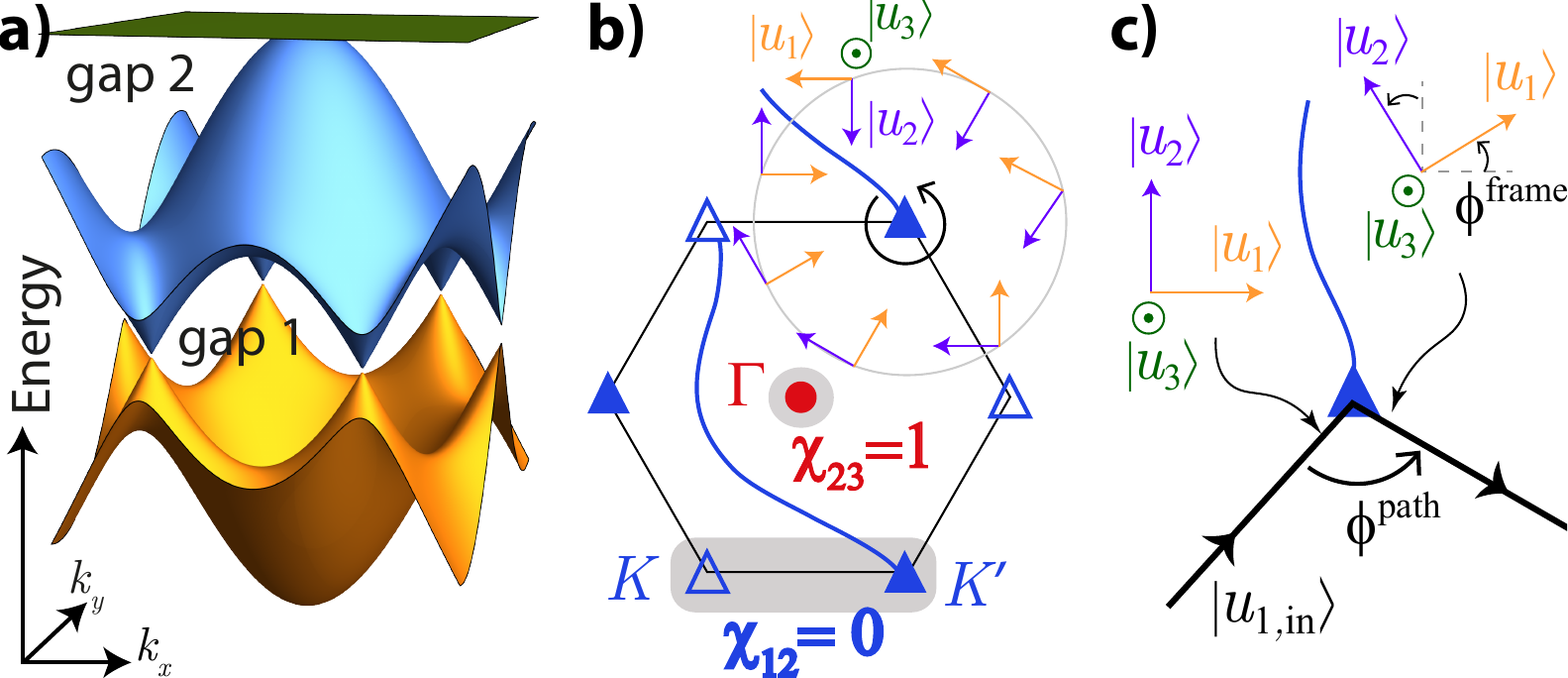}    
\caption{a) Kagome lattice band structure for $\Delta_{\beta}=0$. b) Linear band nodes $(K,K')$ carry opposite non-Abelian frame charges ($\pm i$$\equiv\striangle/\blacktriangle$), and can annihilate ($\chi_{1,2}=0$ in the shaded area). They act as $\pm\pi$ fluxes in gap 1, captured by the DS connecting them.
$\Gamma$ node is formed by two similar-valued frame charges ($+j$$\equiv\bullet$) that are obstructed to annihilate $(\chi_{2,3}=1)$.  The orthonormal frame of the eigenstates (arrows) rotates around a band node, accumulating a $\pi$-phase jump at the DS. c) A wavepacket moving through a node at an angle $\phi^{\mathrm{path}}$, and finishing in a coherent superposition [Eq.~\eqref{eq:windingstates}] due to the frame rotation by an angle $\phi^{\mathrm{frame}}$.  } \label{fig:kagome}
\end{figure}

%%%%%%%%%%%%%%%%%%%%%%%%%%%%%%%%%%%%%%%%%%%%%%%%%%%%%
\parag{Non-Abelian Frame Charges and Euler Class}--- For a real Hamiltonian, the eigenstates $|u_{n}(\bk)\rangle$ at quasimomentum $\bk$ span an orthonormal triad or, in general, vielbein (`frame'):~The space of spectralized Hamiltonians is given by $O(N)/O(1)^N=O(N)/\mathbb{Z}_2^N$ for $n\in N$ bands, where $N>2$. The quotient ensures that flipping the signs of the eigenvectors does not change the Hamiltonian. Band singularities obstruct a unique assignment of the eigenstates and correspond to a $\pi$-rotation between the eigenstates that host the node (see Fig.~\ref{fig:kagome}b)~\footnote{Resulting in the $\pi$-Berry flux of a Dirac cone.}. Namely, the frame rotation ($\phi^{\mathrm{frame}}$) accumulates a $\pi$ Berry phase upon circling around a node, where band singularities act as the analogues of $\pi$ disclination vortices in biaxial nematics~\cite{Wu2019, Beekman20171}. While within the two-band subspace (a single `gap') the nodes act as $\pi$-rotations, we note that $2\pi$ rotations in fact correspond to $-1$ as $\pi_{1}(SO(N))=\mathbb{Z}_2$, accumulated frame charges {\it anticommute} with those in the adjacent gaps. Focusing on three bands, %which can also entail a partial frame~\cite{Peng2022} obtained after projecting, 
the frame charges associated to the nodes $\pi_1(SO(3)/D_2)=Q$ take values in the quaternion group $Q=\{\pm1,\pm i,\pm j, \pm k\}$~\cite{Wu2019}. Consequently, braiding band nodes in momentum space (which can be achieved e.g.~using stress/strain~\cite{Bouhon2020nonabelian,Peng2022}, temperature effects~\cite{chen2021manipulation} or periodic driving~\cite{Slager2024}), converts their charges and can ensure that a specific band subspace hosts similarly-valued charges. In two dimensions, the resulting obstruction to annihilate these charges (nodes) between states $n$ and $(n+1)$ is quantified by the Euler class~\cite{Bouhon2020nonabelian,SI},
\begin{equation} \label{eq:Eulerpatch}
\chi_{n,n+1}{[\mathcal{D}]} = \dfrac{1}{2\pi} \left[\int_{\mathcal{D}}  \mathrm{Eu } ~dk_1\wedge dk_2 - \oint_{\partial \mathcal{D}} \mathcal{A}\cdot d\boldsymbol{k} \right] \in \mathbb{Z},
\end{equation}
evaluated over any patch $\mathcal{D}$ encapsulating these nodes, [Fig.~\ref{fig:kagome}b]. %, measuring the number of pairs that cannot be annihilated due to their similar frame charge. 
Here, we define the Euler form $\mathrm{Eu} = \langle \partial_{k_1} u_n(\boldsymbol{k})\vert \partial_{k_2} u_{n+1}(\boldsymbol{k})\rangle - \langle \partial_{k_2} u_n(\boldsymbol{k})\vert \partial_{k_1} u_{n+1}(\boldsymbol{k})\rangle$ and associated connection one-form, $\mathcal{A} =  \langle u_{n}(\boldsymbol{k})|\boldsymbol{\nabla} u_{n+1}(\boldsymbol{k})\rangle$. 

The braiding and Euler class can effectively be captured by tracking Dirac strings (DSs)~\cite{Slager2024,SI}. These are gauge objects connecting pairs of nodes in each gap (see Fig.~\ref{fig:kagome}) and represent the line across which the sign of the eigenstates change due to the $\pi$-Berry phase induced by the node. Hence, crossing a DS residing in an adjacent gap changes the sign of the frame charge, effectively encoding braiding rules of the band node charges~\cite{SI}.

%%%%%%%%%%%%%%%%%%%%%%%%%%%%%%%%%%%%%%%%%%%%%%%%%%%%%
\parag{Model}---
As a concrete example, we employ a $\mathcal{C}_2 \mathcal{T}$-symmetric Kagome lattice, while our results apply to any system admitting an Euler description. The momentum-space Hamiltonian is written as
\begin{equation}  \label{eq:KagomeH}
    H(\bk) = -2J \sum_{\beta\neq\beta'} \cos{\left(\bk \cdot \bm{d}_{\beta\beta'}\right)} c^{\dagger}_{\beta}c^{\phantom{\dagger}}_{\beta'} + \sum_{\beta} \Delta_{\beta} c^{\dagger}_{\beta}c^{\phantom{\dagger}}_{\beta},
\end{equation}
with nearest-neighbour hopping amplitudes $J$ along the three directions $\bm{d}_{\beta \beta'}$ connecting three sublattices $\beta\in(A,B,C)$ and the annihilation (creation) operator $c^{(\dagger)}_{\beta}$~\cite{Slager2024}. For vanishing sublattice offsets $\Delta_{\beta}$, there are two linear band touching points which carry opposite frame charges (e.g.~$\pm i$, depicted with empty/filled markers in Fig.~\ref{fig:kagome}b), hence a vanishing patch Euler class. The quadratic band touching point in gap 2 harbours $\chi=1$, corresponding to two same-valued frame charges which we can assign to be $+j$~\footnote{The frame-charge assignment to each gap is a gauge choice, where only the relative sign in a gap is physically meaningful and the corresponding Euler class is gauge invariant.}. While quaternion frame charges act as $\pm\pi$ Berry fluxes in a given gap, conventional interferometry methods are insensitive to their signs and, hence, cannot a priori reveal the relative frame charges or their change after braiding. 
%and hence are not capable of revealing any information related to the frame charges or their non-Abelian braiding. We overcome this challenge by treating band nodes in a given gap all together, and construct two interferometric schemes that are capable of distinguishing a total zero or $2\pi$ winding of the frame associated with band degeneracies.

%%%%%%%%%%%%%%%%%%%%%%%%%%%%%%%%%%%%%%%%%%%%%%%%%%%%%
\parag{Method 1:~Interferometry}---
We employ a coherent superposition in the bands forming a singularity (i.e.~the Euler subspace) to deduce the frame rotation around it relative to other nodes in this gap, hence, their relative frame charges.
%\replaced{Consider preparing a wavepacket in momentum space in band 1 $(\ket{u_{1,\mathrm{in}}(\bk)})$, and applying a force to accelerate the wavepacket along a path in momentum space $\bk(t)$. For small enough acceleration and away from any nodes, the motion will be adiabatic, and the wavefunction will acquire only a dynamic phase \footnote{Since the eigenstates are real, there will be no Berry phase. See Supplementary Materials for more discussion about motion through the Brillouin zone \cite{SI}.}. Now suppose that we accelerate the wavepacket into and out of a node at an angle $\phi^{\mathrm{frame}}$ as in Fig.~\ref{fig:kagome}c~\cite{Brown2022}.}{
Consider moving a wavepacket in momentum space in band 1 $(\ket{u_{1,\mathrm{in}}(\bk)})$, which enters and exits a band node at some angle $\phi^{\mathrm{path}}$ as in Fig.~\ref{fig:kagome}c~\cite{Brown2022}. %}
Between the in-going and out-going paths the eigenstates frame is rotated by an angle $\phi^{\mathrm{frame}}$. Therefore, after passing through the band touching point, atoms are in a coherent superposition~\cite{SI}, %The atoms, therefore, are in a coherent superposition after passing through the node (adiabatically with respect to other bands) as,
\begin{align}     \label{eq:windingstates}
    \ket{u_{1,\mathrm{in}}} &\to \ket{u_{1,\mathrm{out}}} \braket{u_{1,\mathrm{out}}|u_{1,\mathrm{in}}}
    +\ket{u_{2,\mathrm{out}}} \braket{u_{2,\mathrm{out}}|u_{1,\mathrm{in}} }  \nonumber   \\
    &=\cos{\phi^{\mathrm{frame}}} \ket{u_{1,\mathrm{out}}} - \sin{\phi^{\mathrm{frame}}} \ket{u_{2,\mathrm{out}}},
\end{align}
given by the overlap with the final eigenstates $\ket{u_{1,\mathrm{out}}}$ and $\ket{u_{2,\mathrm{out}}}$, captured by the frame rotation $\phi^{\mathrm{frame}}$. Note that we here assume an adiabatic motion with respect to the third band, which can be generally satisfied due to the energetic separation~(Fig.~\ref{fig:kagome}a) and the orthonormality of the eigenstates. While the required path precision is within reach of state-of-the-art experiments~\cite{Brown2022}, the desired superposition in the Euler subspace governed by $\phi^{\mathrm{frame}}$ can be still created even if atoms might just miss the node realistically as we show in Supplemental Material (SM)~\cite{SI}. 
%Moreover, %we here consider uniform and symmetric Hamiltonians such as Kagome lattice to demonstrate our scheme, where the frame winds smoothly $\phi^{\mathrm{frame}}=\pm \phi^{\mathrm{path}}/2$ .  
For linear nodes, $\phi^{\mathrm{frame}}$ winds always by $\pi$ as $\phi^{\mathrm{path}}$ winds by $2\pi$. 
To demonstrate our scheme, we consider uniform and symmetric Hamiltonians (e.g.~Kagome) for a smooth winding $\phi^{\mathrm{frame}}=\pm \phi^{\mathrm{path}}/2$, while our methods readily apply to generic nodes~\footnote{In general, $\phi^{\mathrm{frame}}$ can be a generic function of $\phi^{\mathrm{path}}$ but always winds by $\pi$ as the latter winds by $2\pi$. See Ref.~\cite{SI} for more details and examples.}. 

When the system is characterized by an Euler class, the relative phase of the component excited into the second band depends on the chirality of the frame winding, i.e.~the {\it full dreibein} that is naturally encoded by the Hamiltonian evolution.
%Importantly, the sign of the frame-rotation angle $\phi^{\mathrm{frame}}$ relative to the turning angle $\phi^{\mathrm{path}}$ signifies the sign of the non-Abelian frame charge. 
By analyzing two such excitations, we devise an interferometry of the Euler nodes to extract the relative non-Abelian frame charges and, hence, patch Euler class~\eqref{eq:Eulerpatch}. We shall first present the idea in the simple setting of a Kagome model by considering $(K,K')$ nodes, where the $\mathcal{C}_6$ symmetry eliminates complicating effects of dynamic phases, isolating the key physics at play. We subsequently discuss more general settings, including effects of DSs and dynamic phases.

\begin{figure}
\centering
\includegraphics[width=1\linewidth]{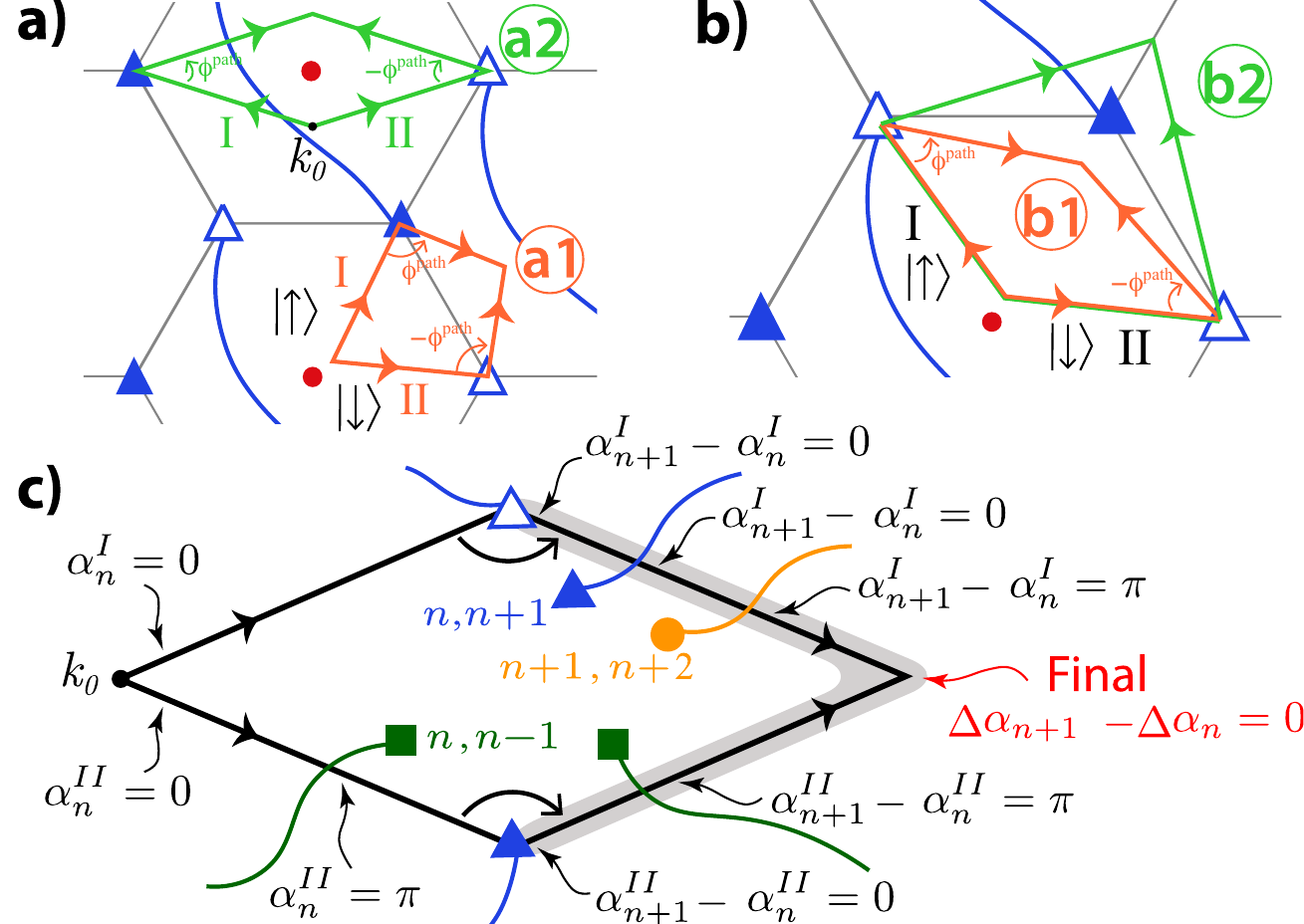}
\caption{ a,b) Interferometry distinguishing relative frame charges of nodes in Kagome lattice. Up- and down-spin atoms are moved along paths I and II, respectively, and recombined. Opposite (a) [same (b)] frame charges result in a single spin [mixture of both spins] as in Table.~\ref{table:kagresults}. In (b2), interferometry paths enclose an additional node, crossing a DS (blue line). c) Illustration of the general interferometry (triangles, circles and squares indicate nodes between different bands as labeled). Final geometric phases $\Delta \alpha_n\! -\! \Delta \alpha_{n+1}\!=\!0$ indicate oppositely charged nodes, which annihilate if brought together within the shaded area, i.e., vanishing patch Euler class. }
\label{fig:InterfrmKgm}
\end{figure}

We consider a wavepacket of a single spin state $\ket{\uparrow}$ localised in band $\ket{u_{1}(\bk_0)}$, starting from $\bk_0$ at equal distance to the targeted Euler nodes. Employing a $\pi/2$-pulse, a superposition of $(\ket{\uparrow}+\ket{\downarrow})/\sqrt{2}$ can be created, where the pseudo-spin can be encoded by, e.g., hyperfine states~\cite{Cooper19_RMP,Duca2015}. As demonstrated in Fig.~\ref{fig:InterfrmKgm}, we split the atoms such that $\ket{\uparrow}$ and $\ket{\downarrow}$ atoms follow symmetric paths I and II, passing through two nodes with angles $\phi^{\mathrm{path}}$ and $-\phi^{\mathrm{path}}$ respectively before recombining. In optical lattices, this may be achieved by using a combination of an applied magnetic field gradient and lattice acceleration~\cite{Duca2015,Cooper19_RMP}.

Assuming that the frame-rotation angle for the node I is $\phi^{\mathrm{frame}}$, the atoms following this path finish in the state $\big(c \ket{u_1} - s \ket{u_2} \big) \ket{\uparrow}$ as in Eq.~\eqref{eq:windingstates}, where $c\equiv \cos{\phi^{\mathrm{frame}}}$ and $s\equiv \sin{\phi^{\mathrm{frame}}}$ and the momentum index is suppressed for simplicity. If the node on path II carries the same charge with no DS in between, the frame features the same chirality on both nodes. However, since the path winds in the opposite direction ($-\phi^{\mathrm{path}}$), the frame-rotation angle experienced on path II is $-\phi^{\mathrm{frame}}$. Conversely, if the nodes have opposite charges, path II corresponds to a frame-rotation angle $\phi^{\mathrm{frame}}$~\footnote{We consider these interferometry angles to be such that the magnitude of frame rotation angle is the same on both paths, which can be controlled e.g.~using the techniques in Ref.~\cite{Brown2022} even for reduced lattice symmetries~\cite{SI}.}.

Considering the paths together, the final state becomes
\begin{equation}
    \frac{1}{\sqrt{2}} \bigg[ \big(c \ket{u_1} - s \ket{u_2} \big) \ket{\uparrow} + \big(c \ket{u_1} \pm s \ket{u_2} \big) \ket{\downarrow}\bigg],
\end{equation}
where $+ (-)$ for down-spins on path II corresponds to same (opposite) charges. Applying another $\pi/2$ pulse before closing the interferometry then yields $c\ket{u_1}\ket{\uparrow} - s \ket{u_2} \ket{\downarrow}$ for similarly-charged nodes, resulting in a mixture of both spin populations, $p_{\uparrow} = \cos^2{\left(\phi^{\mathrm{frame}}\right)}\mathrm{~and~}p_{\downarrow} = \sin^2{\left(\phi^{\mathrm{frame}}\right)}$. %~\footnote{We consider a $\pi/2$ pulse mapping $\ket{\uparrow}\rightarrow(\ket{\uparrow}+\ket{\downarrow})/\sqrt{2}$ and $\ket{\downarrow}\rightarrow(\ket{\uparrow}-\ket{\downarrow})/\sqrt{2}$.}. 
However, for oppositely-charged nodes, the final state $c\ket{u_1}\ket{\uparrow} - s \ket{u_2} \ket{\uparrow}$  consists only of up spins $p_{\uparrow}=1$. 
We demonstrate our interferometry technique along different paths in Fig.~\ref{fig:InterfrmKgm}(a,b) with resulting populations given in Table~\ref{table:kagresults} which are also numerically confirmed, where the relative non-Abelian frame charges are distinguished.

\begin{table}
\begin{tabular}{ |c|c|c|c|c| } 
\hline
 Paths & Final state  & $p_{\uparrow}$ & $p_{\downarrow}$ & Charges\\
 \hline
(a1) & $c \ket{u_1} \ket{\uparrow} - s \ket{u_2} \ket{\uparrow}$ & $1$ & $0$ & opposite \\
\hline
(a2) & $c \ket{u_1} \ket{\uparrow} - s \ket{u_2} \ket{\uparrow}$ &  $1$ & $0$ & opposite \\
\hline
(b1) &$c\ket{u_1} \ket{\uparrow} - s \ket{u_2} \ket{\downarrow}$ &  $\cos^2{\phi^{\mathrm{path}}/2}$ & $\sin^2{\phi^{\mathrm{path}}/2}$ & same \\
\hline
(b2)  &  $c\ket{u_1} \ket{\downarrow} - s \ket{u_2} \ket{\uparrow}$ &  $\sin^2{\phi^{\mathrm{path}}/2}$ & $\cos^2{\phi^{\mathrm{path}}/2}$ & same \\
 \hline
\end{tabular}
\caption{Interferometry results for paths in Fig.~\ref{fig:InterfrmKgm}(a,b). While opposite frame charges result in a single spin species, similar charges yield a mixture. }  \label{table:kagresults}
\end{table}

%%%%%%%%%%%%%%%%%%%%%%%%%%%%%%%%%%%%%%%%%%%%%%%%%%%%%
\parag{General configurations}---
Our interferometry technique readily caters to more complex band structures. The dynamic phases along the two paths considered above (Fig.~\ref{fig:InterfrmKgm}) are equal under $\mathcal{C}_6$ symmetry and cancel in Table~\ref{table:kagresults} (see SM~\cite{SI}). In general when targeted nodes are not at high-symmetry points, dynamic phases must be accounted for. This can be achieved e.g.~by measuring band energies along the interferometry with standard techniques like band mapping~\cite{Greiner01_PRL_phasecoherence,Cooper19_RMP}. We present another method based on performing a reference interferometry loop with twice the acceleration in SM~\cite{SI}.

Furthermore, there can be DSs traversed by the interferometry loop, which change the sign of the relevant eigenstates, leading to an additional $\pi$ geometric phase. While the DS crossed twice in Fig.~\ref{fig:InterfrmKgm}a yields no net effect as in Table.~\ref{table:kagresults}~\footnote{Equivalently, the Dirac string can be deformed past the node along path I and omitted}, the single crossing on path I in Fig.~\ref{fig:InterfrmKgm}b induces a $\pi$ phase shift between the spins and switches $p_{\uparrow}$ and $p_{\downarrow}$, but still entails a spin mixture for the similar-valued frame charges. %Whilst DSs in the Euler subspace don't affect the relative charge, DSs with adjacent bands can.  
We here give a proof for generic DS configurations that our interferometry scheme determines whether the nodes can annihilate if brought together along the paths closing the interferometry loop (Fig.~\ref{fig:InterfrmKgm}c). This reveals the relative frame charges and gauge-invariant Euler class~\eqref{eq:Eulerpatch} within this region. 

We consider two Euler nodes between bands $n$ and $n+1$,  and focus on geometric phases $\alpha_{n(n+1)}^\text{I/II}$ in each band along paths I and II, where dynamic phases can be accounted for as before if required. We now analyze the effect of all possible DSs:
 
%\begin{enumerate}
%\item 
1. Right before the atoms starting in band $n$ enter the nodes, the phase difference between the two paths $\Delta \alpha_n \equiv \alpha_n^\text{II} - \alpha_n^\text{I}$ can be $0$ or $\pi$, depending on any DSs between the initial point and the nodes as depicted in Fig.~\ref{fig:InterfrmKgm}c. These have no effect as it will be carried over to the contributions in both bands on that path below~\cite{SI}.

%\item 
2. Just after exiting the nodes, the phase of the excited component $\alpha^\text{I/II}_{n+1}$ on the two arms depends on the frame charges, where we emphasize that the gauge is naturally fixed by the Hamiltonian evolution. %~\footnote{Namely, the relative $U(1)$ phase of each eigenstate up to DSs is fixed by the Hamiltonian that constitutes evolution operator}. 
If the charges are opposite as in Fig.~\ref{fig:InterfrmKgm}c, the total phase difference between the two paths are the same for both bands $\Delta \alpha_{n+1} - \Delta \alpha_{n}=0$, while there will be a sign change $|\Delta \alpha_{n+1} - \Delta \alpha_{n}|=\pi$ for same charges.

%\item 
3. When the wavepackets are brought together avoiding other nodes, there can be three types of DSs crossed; within the Euler subspace and above/below gap, see Fig.~\ref{fig:InterfrmKgm}c: Crossing e.g.~the yellow DS changes $\ket{u_{n+1}}\rightarrow-\ket{u_{n+1}}$ on path I, which reflects on $\Delta\alpha_{n+1}$, as summarized in Table~\ref{table:DScrossings} for all cases. If the node itself is moved across the yellow DS, its frame charge flips.
%result in changes in $\Delta \alpha_{n(n+1)}$ summarized in Table~\ref{table:DScrossings}. Most importantly, these DSs would also alter the relative sign of the targeted frame charges if they were to be brought together themselves along the closing paths. 
Indeed, we see that each time $\Delta \alpha_{n+1}-\Delta \alpha_n$ changes by $\pi$, the relative charge of the nodes changes. %As a result, if $(\Delta \alpha_n- \Delta \alpha_{n+1}) = 0$, the relative charge has changed an even number of times, and thus the nodes \textit{will} annihilate. If $(\Delta \alpha_n- \Delta \alpha_{n+1}) = \pi$, the nodes have the same charge when brought together and \textit{will not} annihilate.

%\item 
4. The nodes' obstruction to annihilate within a region of the Brillouin zone is captured by the patch Euler class~\eqref{eq:Eulerpatch}. We therefore conclude that the patch Euler class in the area enclosing the second half of the interferometry [i.e.~post-node branches] vanishes $\chi_{n,n+1}=0$ (is finite, $\chi_{n,n+1}=1$) if $|\Delta \alpha_{n+1}-\Delta \alpha_n| = 0\,(\pi)$. Corresponding spin populations (after a $\pi/2$-pulse upon closing the interferometry~\cite{SI}) are given in Table~\ref{table:spintable} which are numerically verified.
%\end{enumerate}

We note that evaluating the Euler class is superfluous for two nodes separated by a reciprocal lattice vector. While we illustrate the relative charges using the simplest Kagome setting [Fig.~\ref{fig:InterfrmKgm}], our scheme applies generically. We validate this by employing a Kagome model with next-nearest-neighbour tunneling terms~\cite{Jiang2021acoustic} which host several more singularities of various charges, see SM~\cite{SI}. Furthermore, under linear periodic driving, this model features an anomalous Euler phase~\cite{Slager2024} where the non-trivial patch invariant arises by virtue of a DS in the anomalous Floquet gap, which we confirm to be captured by our interferometry~\cite{SI}.

\begin{table}
\begin{tabular}{ |c|c|c|} 
\hline
\rule[-10pt]{0pt}{25pt}  DS between bands & Effect on $\Delta \alpha_n$, $\Delta \alpha_{n+1}$ & Relative charge\\
 \hline
$n$ and $n+1$ & Both change by $\pi$ & No change\\
\hline
$n$ and $n-1$ & $\Delta \alpha_n$ by $\pi$ & Changes \\
 \hline
$n+1$ and $n+2$  & $\Delta \alpha_{n+1}$ by $\pi$ &  Changes    \\
 \hline
\end{tabular}
\caption{$\pi$-phase changes acquired by crossing possible DSs in Fig.~\ref{fig:InterfrmKgm}c (shaded area) and corresponding effects on relative frame charges as explained in the text.
%Possible DS crossings (Fig.~\ref{fig:InterfrmKgm}c) after the targeted nodes, and corresponding effects on the phase difference in each band and frame charges if the nodes were to be moved. %(by tuning the Hamiltonian's parameters).
}
\label{table:DScrossings}
\end{table}

\begin{table}
\begin{tabular}{ |c|c|c|c|c| } 
\hline
 $\Delta \alpha_n$ & $\Delta \alpha_{n+1}$ & $p_{\uparrow}$ & $p_{\downarrow}$ & Charges\\
 \hline
$0$ & $0$ & $1$ & $0$ & opposite\\
\hline
$\pi$ & $\pi$ & $0$ & $1$ & opposite\\
\hline
$0$ & $\pi$ & $\cos^2{\phi^{\mathrm{frame}}}$ & $\sin^2{\phi^{\mathrm{frame}}}$ & same\\
\hline
$\pi$ & $0$ & $\sin^2{\phi^{\mathrm{frame}}}$ & $\cos^2{\phi^{\mathrm{frame}}}$ & same\\
 \hline
\end{tabular}
\caption{Effects of possible phase changes [cf.~Table.~\ref{table:DScrossings}] on spin populations in bands $n,\,n+1$ after the interferometry: Relative frame charge depends only on $|\Delta \alpha_{n+1} - \Delta \alpha_n|$).%Spin populations resulting from phase differences between the two paths in bands $n$ and $n+1$.
}
\label{table:spintable}
\end{table}

\begin{figure}
\includegraphics[width=1\linewidth]{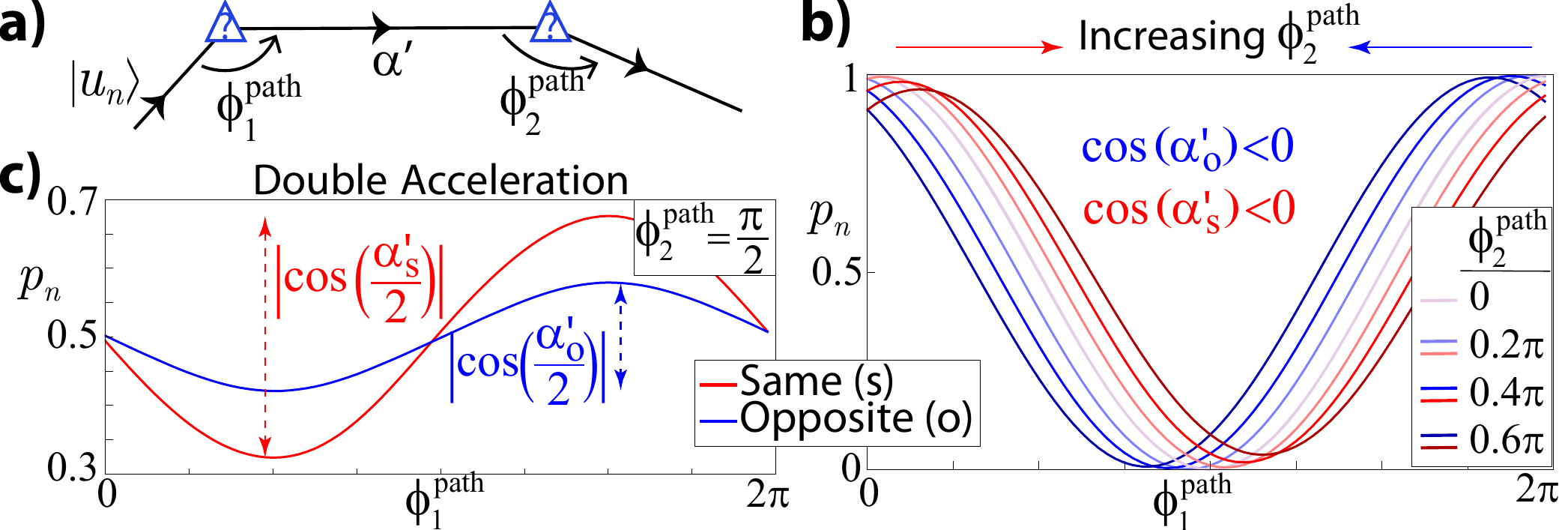}   
\caption{a) Illustration of the consecutive deflection method. %, passing through two nodes with turning angles $\phi^{\mathrm{path}}_1$, $\phi^{\mathrm{path}}_2$. 
Between the nodes, a relative dynamic phase $\alpha'$ is developed. b) Starting from band $n=1$, the final population $p_{1}$ oscillates as a function of $\phi^{\mathrm{path}}_1$. The phase-shift direction for increasing $\phi^{\mathrm{path}}_2$ depends on the relative frame charge and sign of $\cos{\alpha'}$, here demonstrated in Kagome lattice for $(K,K')$ nodes~\eqref{eq:KagomeH}. c) $\cos{\alpha'}<0$ is deduced from Eq.~(\ref{eq:amplitude}) at double acceleration with $\phi^{\mathrm{frame}}_2=\pi/4$. %The phase shift direction is deduced from Eqs.~(\ref{eq:amplitude},\ref{eq:phase}), after determining $\cos{\alpha'}<0$ at double acceleration with $\phi^{\mathrm{path}}_2=\pi/2$.
}  \label{fig:method2}
\end{figure}

%%%%%%%%%%%%%%%%%%%%%%%%%%%%%%%%%%%%%%%%%%%%%%%%%%%%%
\parag{Method 2: Consecutive Deflection}---
Our second method does not require an interferometry loop and offers the advantage of not increasing in complexity when symmetries are removed. This relies on a single wavepacket entering and exiting the two targeted nodes [in the same way as in Eq.~\eqref{eq:windingstates}]  between bands $(n,\,n+1)$ consecutively [Fig.~\ref{fig:method2}a].
Starting in band $n$, the final population $p_n$ in this band depends on the frame-rotation angles $(\phi^{\mathrm{frame}}_1,\phi^{\mathrm{frame}}_2)$, as well as the relative dynamic phase $\alpha'$ developed in between the nodes, which a priori complicates deducing the frame charges~\footnote{Any relative dynamic phase appearing along the pre/post-node sections are irrelevant since they will not affect the populations.}. We circumvent this by expressing it after some algebra~\cite{SI} as $p_n\!\!=\!\!\left[ A \cos{\left(2\phi^{\mathrm{frame}}_1 \!-\!\beta \right)} \!+\! 1 \right]\!/2$, where
\begin{align}   
    A^2 = \cos^2{\left(2\phi^{\mathrm{frame}}_2\right)}+\cos^2{\left(\alpha'\right)}\sin^2{\left(2\phi^{\mathrm{frame}}_2\right)},\label{eq:amplitude} \\
    \tan{\beta}=-\cos{\left(\alpha'\right)} \tan(2\phi^{\mathrm{frame}}_2). \label{eq:phase}
\end{align}
The band population varies with $\phi^{\mathrm{frame}}_1$ sinusoidally with an amplitude $A$ and phase $\beta$. % that depend on $\phi^{\mathrm{frame}}_2$. 
Crucially, the phase shift in Eq.~(\ref{eq:phase}) depends on whether the frame winds in the same/opposite direction around the second node $(\phi^{\mathrm{frame}}_2)$, corresponding to same/opposite frame charges, provided that $\cos{\left(\alpha'\right)}$ is known. The latter can be extracted from the amplitude~\eqref{eq:amplitude}, which reveals only $\cos^2(\alpha')$, by performing the experiment at double the acceleration to halve the dynamic phase. Obtaining $\cos^2{\left(\alpha'/2\right)}$ thus yields $\cos{\alpha'}=2\cos^2{\left(\alpha'/2\right)}-1$. Alternatively, the dynamic phase can be directly calculated from energy measurements~\cite{Greiner01_PRL_phasecoherence,Cooper19_RMP}.

We numerically demonstrate this method by using the $(K,K')$ nodes in Fig.~\ref{fig:method2} (see SM~\cite{SI} for more complicated settings), where $p_{n=1}$ oscillates as a function of the first turning angle $\phi^{\mathrm{path}}_1=2\phi^{\mathrm{frame}}_1$. Repeating for different turning angles in node 2 induces a phase shift, direction of which reveals the relative frame charges. Fig.~\ref{fig:method2}c reveals $\cos\alpha'<0$ in both cases, and therefore the phase $(\beta)$ increases (decreases) for similar (opposite) frame charges.%~\footnote{For $\cos\alpha'>0$, these shifts are reversed}. 

%%%%%%%%%%%%%%%%%%%%%%%%%%%%%%%%%%%%%%%%%%%%%%%%%%%%%
\parag{Discussion}--- 
Our two interferometric schemes for extracting non-Abelian frame charges of real Hamiltonians require control on lattice acceleration and path precision that have been already demonstrated in ultracold atom experiments~\cite{Brown2022,Duca2015}. While the first method involves two measurements--the actual interferometry and in general a reference loop if dynamic phases differ along each path~\cite{SI}--, the second method relies on several measurements for a range of $\phi^{\mathrm{path}}_{1,2}$ values to detect the phase shift but applies independently of underlying lattice symmetries. We note that pseudospins in Method 1 can be employed for guiding atoms along an interferometry loop~\cite{Duca2015} and further combining with band specific readouts of atoms reveals individual terms in Table~\ref{table:kagresults}. Method 2 can be fine-tuned for a specific experiment as well. For example, for known frame windings around individual nodes (e.g.~uniform under $C_6$ symmetry~\cite{SI}),
one requires only three measurements: Two for observing $p_n$ under double acceleration by choosing $|\phi^{\mathrm{frame}}_1|=|\phi^{\mathrm{frame}}_2|=\pi/4$ and reversing the sign of $\phi^{\mathrm{frame}}_1$, which yield
$\sin^2{(\alpha'/4)}$ and $1-\sin^2{(\alpha'/4)}$ to find $\cos^2{(\alpha'/2)}$. Repeating the first measurement at normal acceleration then discloses $1-\cos^2{(\alpha'/2)}~\,(\cos^2{(\alpha'/2)})$ for similar (opposite) charges.
We further demonstrate our techniques' robustness under experimental imperfections in~\cite{SI}, putting them within reach of state-of-the-art experiments~\cite{Duca2015,Brown2022}.

The interferometry and consecutive deflection methods reveal the relative frame charge along the path joining the nodes. Hence, by varying the route connecting the nodes, our techniques pave the way for directly demonstrating the non-Abelian and path dependent nature of braiding in experiments. Given the essential role of such charges in constituting Euler class~\cite{Ahn2019nielsen, Bouhon2020nonabelian, bouhon2020geometric}, these serve as a crucial route towards experimentally probing multi-gap topologies in and out of equilibrium~\cite{Slager2024}. % encountered in atomic simulators.  
%Distinguishing such charges in our outlined procedures also gives impetus for analyzing in other phases beyond these settings.

%%TC:ignore
\begin{acknowledgments}
	{\it Acknowledgments---}
O.B.~thanks James Walkling and Calvin Hooper for productive discussions. We acknowledge discussions with Dan Stamper-Kurn and Adrien Bouhon. O.B.~acknowledges funding from the Clarendon Fund, Merton College, and a Leverhulme Studentship. R.-J.~S.~acknowledges funding from a New Investigator Award, EPSRC grant EP/W00187X/1, a EPSRC ERC underwrite grant  EP/X025829/1, and a Royal Society exchange grant IES/R1/221060 as well as Trinity College, Cambridge.  F.N.\"U.~acknowledges funding from the Marie Sk{\l}odowska-Curie programme of the European Commission Grant No 893915, Simons Investigator Award [Grant No.~511029], Trinity College Cambridge, and thanks the Aspen Center for Physics for their hospitality, where this work was partially funded by a grant from the Sloan Foundation. 
\end{acknowledgments}

\bibliography{references}

\begin{thebibliography}{56}
\bibitem [{SI()}]{SI}%
  \BibitemOpen
  \href@noop {} {}\bibinfo {note} {See {S}upplemental {M}aterial ({SM}) for
  more details on Sec I: Frame charges and Dirac strings; Sec II: Moving
  through the BZ under a force and through a node where the motion is
  non-adiabatic, analysing also effects of experimental imperfections; Sec III:
  Method 1 and detailed accounting of dynamic phases; Sec IV: Method 2; and Sec
  V: Further numerical calculations and examples in Kagome lattice with
  next-nearest-neighbor tunneling terms and anomalous Euler phases under
  periodic driving.}\BibitemShut {Stop}%
\end{thebibliography}

\clearpage
\onecolumngrid

\appendix*

\setcounter{equation}{0}
\setcounter{page}{1}
\makeatletter
\renewcommand{\theequation}{S\arabic{equation}}
\renewcommand{\thefigure}{S\arabic{figure}}
\renewcommand{\thetable}{S\arabic{table}}

\section*{Supplemental Material for ``Interferometry of non-Abelian band singularities and Euler class topology"}

\subsection{I. Frame Windings, Dirac Strings, and Non-Abelian Charges}

Frame charges, Dirac strings, and their non-Abelian properties are described in-depth in Ref.~\cite{Bouhon2020nonabelian}. Here we provide a brief exposition to the key concepts. 

In $\mathcal{C}_2 \mathcal{T}$ symmetric systems, the eigenvectors are real and orthogonal, and thus at each quasimomentum form an orthonormal frame $\{\ket{u_n(k)}\}_{n=1}^N$ \cite{Bouhon2020nonabelian, Wu1273}. However, each eigenvector is only defined up to an overall sign $\pm$ (a gauge degree of freedom). Considering a closed path, since the initial and final Hamiltonians will be identical, the accumlated angle defines a `frame charge' associated with the path. In particular, a frame charge may be nontrivial when nodes are enclosed. For example, for a path enclosing a single linear node between two bands (the `principal bands'), the eigenvectors corresponding to the principal bands undergo a $\pi$ rotation as we circle the node \cite{Bouhon2020nonabelian}.  Such a $\pi$ winding means that it is not possible to define a globally smooth gauge and there is a line across which the gauge switches so the sign of these two bands changes. This line, similar to a branch cut, is called a Dirac string \cite{Ahn2019nielsen}, and is demonstrated in Fig.~1(b). A Dirac string may either terminate on a node (joining a pair of nodes) or may cross the entire Brillouin zone (e.g.~as in the anomalous Floquet phases introduced in Ref.~\cite{Slager2024}) after nodes are recombined over the Brillouin zone edge.

We can see this winding more explicitly as follows. For a linear node, expanding in wavevectors about the degenerate point in the subspace corresponding to these bands, the Hamiltonian must take the form \cite{Bouhon2020nonabelian}
\begin{equation}
    H_2 = \sum_{i=1}^2 \sum_{j=0}^3 k_i h_{ij} \sigma_j,
\end{equation}
where the coefficients $h_{ij}$ are real and $h_{i2}=0$. Neglecting the identity component (which does not affect the eigenstates),
\begin{equation}
    H_2=\begin{pmatrix}
        k_i h_{i3} & k_i h_{i1} \\  k_i h_{i1} & -k_i h_{i3} 
    \end{pmatrix}.
\end{equation}
Making the linear change of basis
\begin{equation}
\label{transformation}
    \begin{pmatrix}
        \kappa_1 \\ \kappa_2
    \end{pmatrix}=\begin{pmatrix}
        h_{13} & h_{23} \\  h_{11} & h_{21} 
    \end{pmatrix} 
    \begin{pmatrix}
        k_1 \\ k_2
    \end{pmatrix},
\end{equation}
we finally have 
\begin{equation}
     H_2=\begin{pmatrix}
        \kappa_1  &  \kappa_2 \\ \kappa_2 & -\kappa_1
    \end{pmatrix}=|\kappa| \left[ \cos{\theta} \sigma_1 +\sin{\theta}\sigma_3\right],
\end{equation}
The eigenstates take the form
\begin{equation}
    \ket{u_1} = \begin{pmatrix}
  \cos{\theta/2} \\
  -\sin{\theta/2}\end{pmatrix}, 
  \ket{u_2} = \begin{pmatrix}
  \sin{\theta/2} \\
  \cos{\theta/2},
\end{pmatrix}
\end{equation}
i.e. the frame angle is $\phi^{\mathrm{frame}}=\theta/2$. As we move around the node $\theta$ varies from $0$ to $2\pi$, and therefore the frames wind by an angle of $\pi$. %The direction of winding corresponds to the sign of the determinant of the change of basis.
 This winding is demonstrated in Fig.~1b in the main text. The change of basis in Eqn.~(\ref{transformation}) is not necessarily orthogonal, and thus a uniform winding of $\theta$ here does not always correspond to a uniform winding of the frames in terms of the angle in the original $(k_1,k_2)$ basis.

\begin{figure*}
\includegraphics[width=0.7\linewidth]{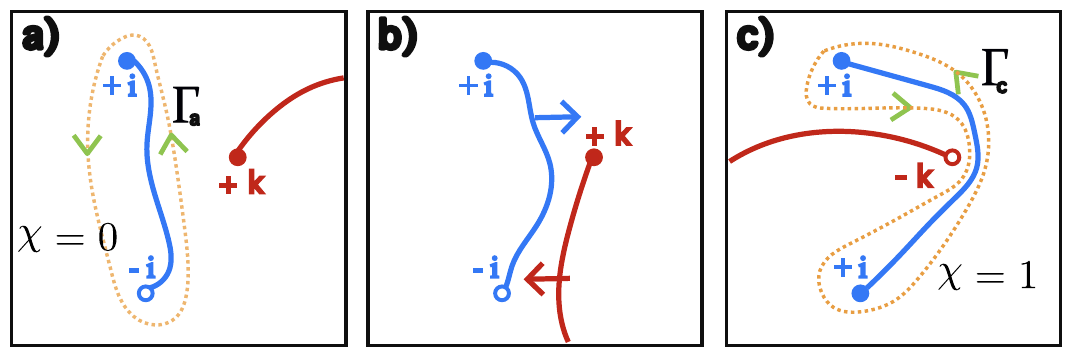}
\caption{Demonstration of how the relative charge of two nodes depends on the region considered. Nodes and Dirac strings between bands 1 and 2 are in blue and labelled by $i$, those between bands 2 and 3 are in red and labelled by $k$. a) In the region $\Gamma_a$ the nodes have opposite charge, with a patch Euler class $\chi=0$, or equivalently the winding of the frame along $\Gamma_a$ is $0$. b) Pulling the Dirac strings through the node in the other gap change their charge, resulting in a smoothly defined gauge within $\Gamma_c$. c) In the region $\Gamma_c$ the nodes have the same charge and thus $\chi=1$. The frame winds by $2\pi$ along $\Gamma_c$ as a manifestation of the non-Abelian nature of the band nodes. We emphasise that each subfigure is the same physical situation, but each with a different gauge choice. }
\label{nodepaths}
\end{figure*}

The sign of (the quaternion charge of) the node refers to the direction in which the eigenvectors wind. While the sign of a given node can be changed through a gauge transformation, %, and thus the charge can change. This means that the charge of an Euler node is not absolutely defined, in contrast to the chirality of a Weyl node \cite{Vanderbilt2018}. Although absolute charge cannot be defined, 
relative charge in a band subspace can be defined by smoothly fixing the gauge between the nodes. For two similarly-charged nodes in a given gap, the net frame winding around a path encircling them is $2\pi$, indicating an obstruction to the ability of these nodes to annihilate. For oppositely charged nodes, the net winding is $0$, and therefore the nodes may annihilate. The relative frame charges can be path dependent as demonstrated in Fig.~\ref{nodepaths}, in which two nodes between the principal bands ($\ket{u_1}, \ket{u_2}$) are labelled by $\pm i$, and one `adjacent' node by $+k$. %where the $\pm$ indicates charges and the letter indicates which gap the node is in.
Suppose that within the enclosed region in (a), the two $i$ nodes have opposite charge. We now ask what the relative charge would be for the region on the other side of the $k$ node, as in (c). In fixing the gauge to be smooth within that region, we must move the red Dirac string outside of this region. Pulling this string through the $-i$ node reverses the signs of $\ket{u_2}, \ket{u_3}$, and thus the winding of the $\ket{u_1},\ket{u_2}$ frame changes. As a result, within this region in (c) the $i$ nodes have the same charge. We stress that the relative stability is a gauge invariant physical quantity and even motivates a topological invariant, the patch Euler class. Indeed, considering region $\Gamma_a$ in Fig.~\ref{nodepaths}c the $i$ nodes still can annihilate in this patch as one transverses the Dirac string of the adjacent gap, a similar analysis holds for considering region $\Gamma_c$ in Fig.~\ref{nodepaths}a. This motivates a puzzle approach~\cite{Peng2022,Peng2022Multi} in which one fixes the gauge at a patch by referring to relative stability, e.g.~on region $\Gamma_a$ and build up the global topological configuration using that the frame charge changes sign when crossing the Dirac strings residing in adjacent gaps as well as combination rules for Dirac strings~\cite{Slager2024}.

Another closely-related perspective is that of braiding, see Refs.~\cite{Bouhon2020nonabelian,Slager2024}. By tuning parameters of the Hamiltonian, the positions of nodes can be adjusted. Moving a principal node around an adjacent node changes the charges of the nodes, which can be interpreted as due to each node crossing the Dirac string associated with the other. %Within this perspective, we can consider bringing nodes together: 
If two nodes in a given gap have opposite charge, they can annihilate; but if the nodes have the same charge they cannot, and will typically `bounce' \cite{Bouhon2020nonabelian}. Due to the braiding, this ability to annihilate will be path dependent. If the nodes are brought together within $\Gamma_a$ in Fig.~\ref{nodepaths} they can annihilate, but within $\Gamma_c$ they cannot. 

The above labelling of the nodes with quaternion charges $i,j,k$ is not unique. %as charges can be relabeled by changing the gauge and assignment of frame. 
Consider nodes only in a 3-band subspace, and suppose we circle a node between bands 1-2 and another between bands 2-3. Composing a $\pi$ rotation of $\ket{u_1},\ket{u_2}$ followed by a $\pi$ rotation of $\ket{u_2},\ket{u_3}$, is in fact equivalent to a $\pi$ rotation of $\ket{u_3},\ket{u_1}$~\cite{Slager2024}. Considering different pairs of bands to get equivalent relations, and noting that a trivial or $2\pi$ winding in any gap is equivalent, we can see that the set of charges can be described by the quaternion group with charge composition described by quaternion multiplication rules. More explicitly, we define the charges of nodes e.g.~of bands 1-2 as $\pm i$ (sign reflecting the opposite direction of winding), bands 2-3 as $\pm k$, bands 3-1 as $\pm j$, the trivial winding as 1 and the $2\pi$ winding as -1. Then charge composition is reflected by
\begin{equation}
\label{quaternions}
    i^2=j^2=k^2=-1,~ij=k=-ji,
\end{equation}
since $\pi_1(SO(3)/D_2)= Q$.
We emphasise that both the assignment of $i,j,k$ to each gap is a gauge choice, in addition to defining the signs of a given charge, but their relative charge when compared with a consistent gauge is a physically meaningful quantity.
%This non-Abelian behaviour, which can also be interpreted as charges taking on quarternion values as described in Appendix~\ref{quarternions}, is a striking aspect of multigap topology. 
The ability to annihilate within a region is captured by the patch Euler class over the region $\mathcal{D}$
\begin{equation} \label{eq:SI:Eulerpatch}
\chi_{n,n+1}{[\mathcal{D}]} = \dfrac{1}{2\pi} \left[\int_{\mathcal{D}}  \mathrm{Eu } ~dk_1\wedge dk_2 - \oint_{\partial \mathcal{D}} \mathcal{A}\cdot d\boldsymbol{k} \right] \in \mathbb{Z},
\end{equation}
where the Euler form $\mathrm{Eu} = \langle \partial_{k_1} u_n(\boldsymbol{k})\vert \partial_{k_2} u_{n+1}(\boldsymbol{k})\rangle - \langle \partial_{k_2} u_n(\boldsymbol{k})\vert \partial_{k_1} u_{n+1}(\boldsymbol{k})\rangle$ and associated connection one-form $\mathcal{A} =  \langle u_{n}(\boldsymbol{k})|\boldsymbol{\nabla} u_{n+1}(\boldsymbol{k})\rangle$. %For a region containing two nodes, if the nodes have opposite charge within that region (and therefore can annihilate) the patch Euler class is $\chi=0$, whereas if the nodes have the same charge (and therefore cannot annihilate) the patch Euler class has magnitude $\chi=1$. The dependence of patch Euler class on region is demonstrated in Fig.~\ref{nodepaths}.

\subsection{II. Motion through the Brillouin Zone}

\begin{figure}
\includegraphics[width=0.75\linewidth]{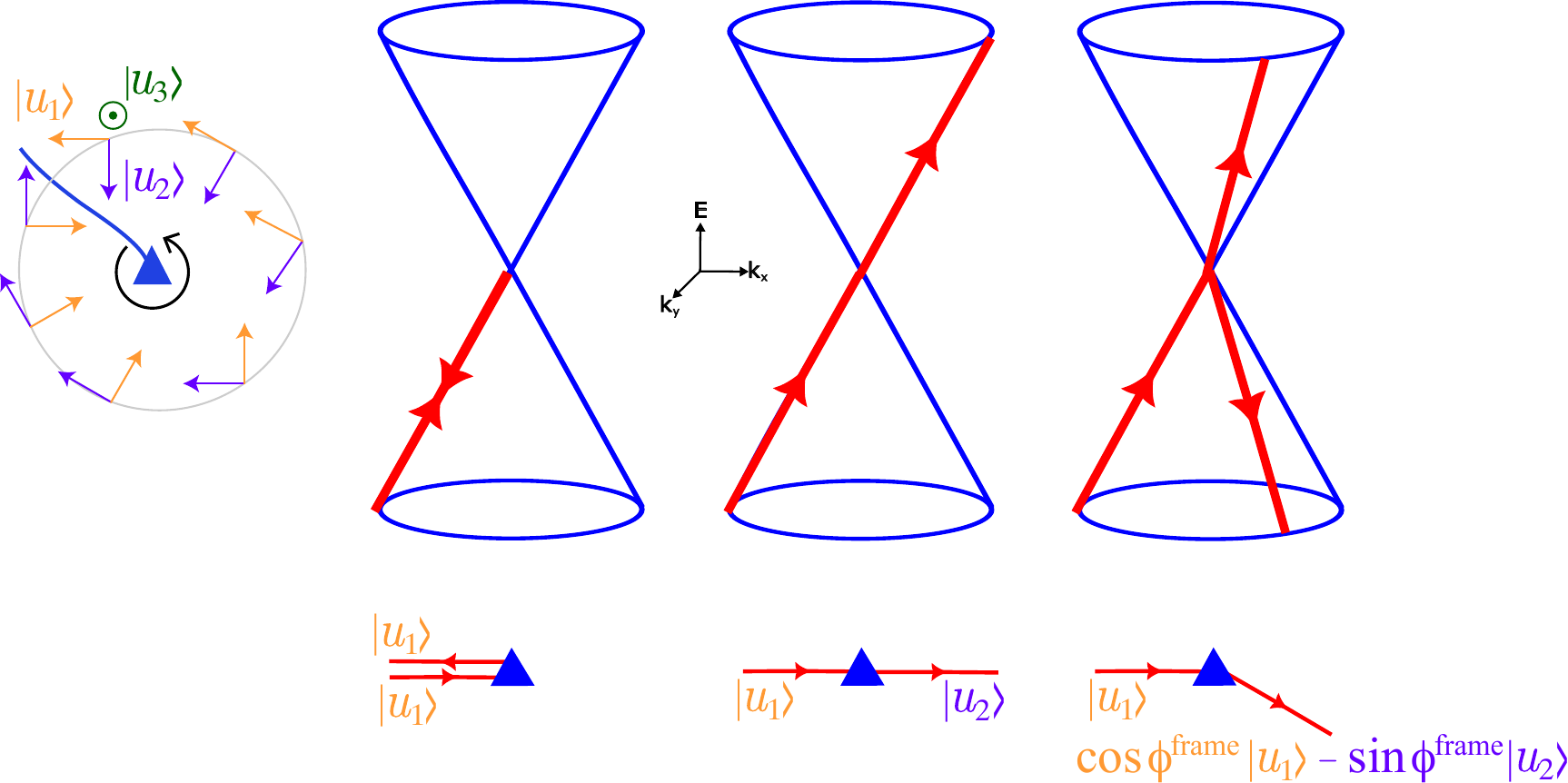}
\caption{A schematic demonstrating the occupation of eigenstates as we enter and exit a linear node at $\phi^{\mathrm{frame}}=0$, $\pi$, and some intermediate angle.  On the left we repeat the demonstration of frame winding from the main text; the eigenstates corresponding to the node rotate by $\pi$ as we circle the node, resulting in the final band populations indicated on and below the cones.}
\label{nodeevol}
\end{figure}

We emphasise that while moving the wavepacket through the BZ and for the resulting superposition given in Eq.(3) of the main text, the motion will be adiabatic for small enough acceleration and away from any nodes. This adiabaticity with respect to the third band, or all other bands outside of the Euler subspace can be in general easily satisfied as these bands will be energetically separated. We note that also within the Euler subspace, no excitations occur until the band touching point due to the orthonormality of the frame eigenstates along the in-going path and the wavefunction will acquire only a dynamic phase as will be shown below. Since the eigenstates are real, there will be no Berry phase acquired along these straight paths. 

In this section we now give a detailed discussion of the dynamics of a wavepacket under an applied force, before applying this to the motion in and out of a node. We discuss the roles of adiabaticity and the effect of `missing' the node by a small amount. Several of these concepts are also discussed in the Supplementary Material for Ref.~\cite{Brown2022}.

\subsubsection{Motion under an applied force}

When an external force $\bm{F}$ is applied to particle which otherwise has Hamiltonian $H$, the Schrödinger equation is
\begin{equation}
\label{schrodinger}
    i \partial_t \ket{\psi} = (H- \bm{F} \cdot \bm{r}) \ket{\psi}.
\end{equation}
We now seek to write this in the basis of Bloch wavefunctions. Suppose the initial state has definite (quasi) momentum $\bk$
\begin{equation}
    \ket{\psi}=e^{i \bk\cdot \bm{r}} \ket{u(\bk)},
\end{equation}
where $\ket{u(\bk)}$ is a superposition of bands
\begin{equation}
    \ket{u_k}=\sum_n c_n \ket{u_{n}(\bk)}.
\end{equation}
The effect of the force is that $\dot{\bk}=\bm{F}$ \cite{Chong2010}, such that through an applied force we can choose the path through momentum-space $\bk(t)$. Therefore, the wavefunction will evolve as
\begin{equation}
    \ket{\psi(t)}= e^{i \bk(t) \cdot \bm{r}} \ket{u(t)},
\end{equation}
with 
\begin{equation}
    \ket{u(t)} = \sum_n c_n (t) \ket{u_n(\bk(t))}.
\end{equation}

Substituting this into Eqn.~(\ref{schrodinger}), we obtain a form of Schrödinger's equation which acts within the Hilbert space of Bloch states
\begin{equation}
\label{schrodingerbloch}
    i \partial_t \ket{u (t)} = H_{\bk(t)} \ket{u(t)},
\end{equation}
where 
\begin{equation}
    H_{\bk} = e^{-i \bk \cdot \bm{r}} H e^{i \bk \cdot \bm{r}}
\end{equation}
is the Bloch Hamiltonian, and $\bk(t)$ now acts as a varying external parameter yielding a time-dependent Hamiltonian $H_{k(t)}$.

If we now initialise the system in a band $\ket{u_{n,k(0)}}$ at wavevector $\bk (0)$, and apply a sufficiently small force that the motion is adiabatic, then the system remains in the same band and accumulates a dynamic phase $\alpha' = \int E_n(k(t))~dt$, with
\begin{equation}
    \ket{u(t)}=e^{-i\int E_n(k(t))~dt}\ket{u_{n,k(t)}}.
\end{equation}
Since the bands are real, no Berry phase is acquired. When a node is encountered and the motion is no longer adiabatic, transitions between the bands can occur; this is explored in the next section.

\subsubsection{Motion through a node}

Consider a single Dirac cone of infinite size and uniform winding, with Hamiltonian $H=v(k_{x}\sigma_{x}+k_{y}\sigma_{z})$. Around the node, the eigenstates take the form 
\begin{equation}
    \ket{u_1(\phi^{\mathrm{path}})}=\begin{pmatrix}
        \sin{\phi^{\mathrm{path}}/2} \\
        -\cos{\phi^{\mathrm{path}}/2}
    \end{pmatrix},~
    \ket{u_2(\phi^{\mathrm{path}})}=\begin{pmatrix}
        \cos{\phi^{\mathrm{path}}/2} \\
        \sin{\phi^{\mathrm{path}}/2}
    \end{pmatrix}.
\end{equation}
As we wind round the node once, taking $\phi^{\mathrm{path}}$ from $0\to2\pi$, the frame angle $\phi^{\mathrm{frame}}=\phi^{\mathrm{path}}/2$ winds from $0\to\pi$. 

Suppose that we take a path which enters and leaves the node at an angle $\phi^{\mathrm{path}}=2\phi^{\mathrm{frame}}$, corresponding to 
\begin{equation}
\boldsymbol{k}(t)
    =
    \begin{pmatrix}
        |\beta t|\cos{\phi^{\mathrm{frame}}} \\
        \beta t\sin{\phi^{\mathrm{frame}}} 
    \end{pmatrix},
\end{equation}
where we have expressed this in terms of $\phi^{\mathrm{frame}}$ such that our following discussion also applies to nodes with non-uniform winding. 

For $t<0$ and separately for $t>0$, the eigenstates are constant. As a result, independently of the acceleration $\beta$, excitations only occur at the node $t=0$. Starting in band $\ket{u_1}$ at time $t_0<0$, the state evolves as 
\begin{equation}
    \ket{u(t)}=\begin{cases}
        e^{iv(t^2-t_0^2)/2}\ket{u_1(-\phi^{\mathrm{frame}})} & t<0 \\
        e^{ivt^2/2}\cos{\phi^{\mathrm{frame}}}\ket{u_1(\phi^{\mathrm{frame}})}+e^{-ivt^2/2}\sin{\phi^{\mathrm{frame}}}\ket{u_2(\phi^{\mathrm{frame}})} & t>0.
    \end{cases}
\end{equation}
To understand this, notice that for $t<0$ we stay in an unchanging eigenstate $\ket{u_1(-\phi^{\mathrm{frame}})}$ of the Hamiltonian, so time evolution only acquires a phase. For $t>0$, the eigenstates are now $\ket{u_{1/2}(+\phi^{\mathrm{frame}})}$, and therefore we decompose our $t=0$ state in terms of the new eigenstates $\ket{u_1(-\phi^{\mathrm{frame}})} = \cos{\phi^{\mathrm{frame}}}\ket{u_1(\phi^{\mathrm{frame}})}-\sin{\phi^{\mathrm{frame}}}\ket{u_2(\phi^{\mathrm{frame}})}$, before applying their respective dynamic phase factors.

Before moving on, let's note some important points. First, since the bands eigenstates are always real, the $\pi$ Berry phase upon circling around the node and the $\pi$ Berry flux of the node is captured by the Dirac string, with our eigenstates picking up a minus sign as we wind once. Secondly, the acceleration $\beta$ here was irrelevant, due to the scale-invariance of the Dirac Hamiltonian and path. A more realistic picture will be discussed in the following section.

\subsubsection{Non-adiabatic motion through a node}

\begin{figure*}
\includegraphics[width=0.7\linewidth]{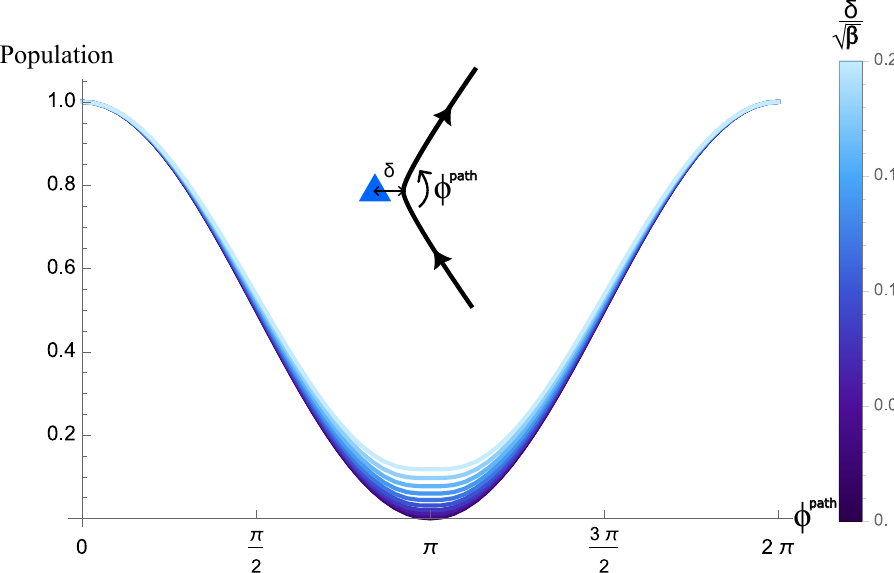}
\caption{Population which remaining in the initial band after missing the node by some amount $\delta$, with an acceleration $\beta$. As long as $\delta$ is sufficiently small ($\delta^2/\beta \ll 1$), the populations and dynamic phases lie close to the ideal case.}
\label{s6}
\end{figure*}

In this section we consider what happens when our path `misses' the node by a small amount, and demonstrate that for sufficiently small misses the results of the previous section still hold. We shall continue to study the same single Dirac cone as in the previous section, but modify the path slightly to 
% \begin{equation}
% \boldsymbol{k}(t)
%     =
%     \begin{pmatrix}
%         |\beta t|\cos{\phi^{\mathrm{frame}}}+\delta \\
%         \beta t\sin{\phi^{\mathrm{frame}}} 
%     \end{pmatrix},
% \end{equation}
\begin{equation}
\boldsymbol{k}(t) = \begin{pmatrix}
        k_x \\
        k_y
    \end{pmatrix}
    =
    \begin{pmatrix}
        \sqrt{(\beta t\cos{\phi^{\mathrm{frame}}})^2 + \delta^2} \\
        \beta t\sin{\phi^{\mathrm{frame}}} 
    \end{pmatrix}
\end{equation}
where we now miss the node by an amount $\delta$ as illustrated in Fig.~\ref{s6}.

% \note{I'm pretty sure a rigorous statement is that the transition probability is $\sin^2{\phi} > p > e^{-\pi \delta^2/\beta} \sin^2{\phi}$, which would be really nice to prove as opposed to just stating the condition below... I've checked this as much as possible numerically but haven't yet found a way to do it analytically. }

Provided that $v\frac{\delta^2}{\beta}\ll 1$, the populations (and dynamic phases - since the difference in dynamic phase will be roughly the change in energy $v\delta$ over a time period $\frac{\delta}{\beta}$) shall be close to those for $\delta=0$, adiabatically. This is demonstrated in Fig.~\ref{s6}, in which the final population in the initial band is plotted as a function of angle $\phi^{\mathrm{frame}}$ for a range of $\delta$. 

$\beta$ is itself bounded by the constraint that motion is fully adiabatic away from the node. This depends on the exact band structure in question, through the energy gaps between the nodes, the rate at which eigenvectors change, and the extent of the node. The extent of the node is simply the region over which the linear expansion in wavevector about the band touching point is valid, such that the rate of winding is fixed. Numerically we find that, for all nodes studied, the extent of the node is at least $\delta k \sim 0.1$. Intuitively, we require perfectly adiabatic motion outside of this region, but inside this circle we are trying to `project' the wavefunction into the bands at some angle around the node, and therefore within this region non-adiabaticity is allowed (and indeed required for non-zero $\delta$ through $v\delta^2/\beta \ll 1$). 
% The condition that we are adiabatic away from this region provides a strong constraint: numerically we find that $\beta \lesssim 0.2$, in order to maintain adiabaticity to within $1\%$. This condition further maintains adiabaticity with the 3rd band at all times. Of the nodes considered in the text, the largest gradient is $v=0.86$ for the Kagome nodes, which places the requirement $\delta\ll 0.5$, which appears to lenient compared to the extent of the node. 

% For example, considering the K and K' points of Kagome along an asymmetric path, with $\beta=0.3$ (as the single acceleration) and taking $\delta=0.1$ for both nodes, we find $\mathrm{abs}(x-(2y-1)^2) + \mathrm{abs}(y-(2z-1)^2)=0.01$ in band 1 and $=0.03$ in band 2. Using the same parameters to study K and K points we get $0.02$ in band 1 and $1.45$ in band 2. Tuning $\beta$ to try to find the worst case scenario brings us down to $0.9$ in band 2, which is still clearly distinguishable from 0. This results indicate that even once further experimental errors (such as in band and spin population measurements, or in the ability to double/quadruple the acceleration accurately) are accounted for, the relative charge of the two nodes should be clearly measurable. 

Further constraints arise from practical considerations. First, the wavepacket possesses finite spread in momentum space. For our methods to work, we require that the phase of the wavefunction is approximately the same over the entire wavepacket. The range of dynamic phases can be estimated from the range of average energies multiplied by the total time taken for the interferometry. In general this condition will be similar to the above condition, as $\frac{\delta k^2}{\beta}\ll 1$. Second, the lattice has finite size (typically 50x50). This is equivalent to a momentum space spreading of order $0.02$. Since this is within all previous bounds, it should not have a significant affect. 

In Section~V, we numerically confirm that our techniques are resilient to such physical considerations.
We also point out that Ref.~\cite{Brown2022} experimentally demonstrates that such paths which slightly `miss' the Dirac point, and gap openings due to effects such as the vector a.c. Stark shift, do not significantly affect the physics. Further, they explicitly demonstrate the coherence evolution of the wavevector after it has passed through one node, which is also crucial to our techniques.

\subsection{III. Details for Method 1: Interferometry}
\label{m1details}

Here, we demonstrate explicitly how the wavefunction evolves along the interferometry paths and give the general form of the final wavefunction. We then describe in detail how the dynamic phases may be accounted for, to extract information about the geometric phases.

\begin{figure}
\includegraphics[width=1\linewidth]{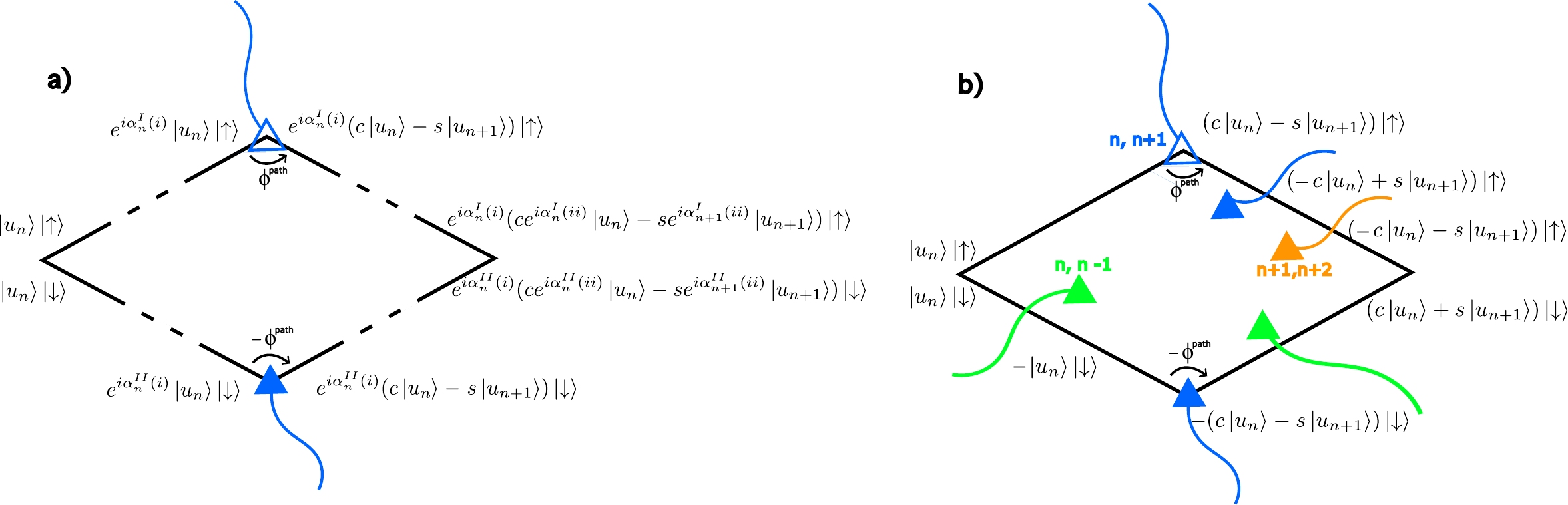}
\caption{ Evolution of the wavefunction along the paths taken during interferometry (note that in the gauge chosen the nodes have opposite charge). a) A general example demonstrating how phases accumulate, where dashed lines indicate that there may be some set of complicated Dirac string crossings. In particular, notice that the phases accumulated by band $n$ in the first half of the path (labelled by (i) for the phases) carries over to the excited component.  b) A specific example configuration of DSs. In this case, $\Delta \alpha_{n+1}-\Delta \alpha_{n}=0$, indicating opposite charges if they were to be brought together along the closing paths. In (b), dynamic phases are not included.}
\label{wavefuncevol}
\end{figure}

Along path I, the wavefunction evolves into
\begin{equation}
    \left(c e^{i\alpha_n^I}\ket{u_n} -s e^{i\alpha_{n+1}^{I}} \ket{u_{n+1}}\right) \ket{\uparrow},
\end{equation}
where $c\equiv \cos{\phi^{\mathrm{frame}}_I},s\equiv \sin{\phi^{\mathrm{frame}}_{I}}$ and $\phi^{\mathrm{frame}}_I$ is the angle by which the frame rotates between ingoing and outgoing paths (see Fig.~1c). The phase $\alpha_n^I$ consists of dynamic and geometric contributions (which are changes by either $0$ or $\pi$) and from the pre- (i) and post-node (ii) sections of the path, $\alpha_n^I=\alpha_n^I(i)+\alpha_n^I(ii)$, as demonstrated in Fig.~\ref{wavefuncevol}. $\alpha_{n+1}^I$ consists of the phase developed in band $n$ before the node, and the dynamic/geometric phases accumulated in band $n+1$ after the node, $\alpha_{n+1}^I=\alpha_n^I(i)+\alpha_{n+1}^I(ii)$. Along path II we similarly have the final wavefunction
\begin{equation}
    \left(c e^{i\alpha_n^{II}}\ket{u_n} -s e^{i\alpha_{n+1}^{II}} \ket{u_{n_1}}\right) \ket{\downarrow}.
\end{equation}
Here, the relative directions of frame winding $\phi^{\mathrm{frame}}_{II}=\pm\phi^{\mathrm{frame}}_I$ leads to a relative sign in the component excited at the node. %(i.e.~demonstrated explicitly as $\mp$ in Fig.~\ref{wavefuncevol}(a)). 
Different than the main text, we can here assume that this frame winding is encompassed in the final phase $\alpha^{II}_{n+1}$ for compactness. It is also worth noting here that, in practice, the paths are brought together by applying a $\pi$-pulse at the nodes, which switches the spin indices. However, this is just equivalent to a relabelling of the paths, and so don't consider this in the main text for simplicity.

Applying a $\pi/2$-pulse which maps $\ket{\uparrow} \to \frac{1}{\sqrt{2}} (\ket{\uparrow} + \ket{\downarrow})$ and $\ket{\downarrow} \to \frac{1}{\sqrt{2}} (\ket{\uparrow} - \ket{\downarrow})$, and upon recombining the wavepackets we obtain
\begin{equation}
    \frac{1}{2}\ket{\uparrow} \left[ c\left(e^{i\alpha_n^I} + e^{i\alpha_n^{II}} \right) \ket{u_n} - s\left(e^{i\alpha_{n+1}^I} + e^{i\alpha_{n+1}^{II}} \right) \ket{u_{n+1}} \right] + \frac{1}{2}\ket{\downarrow}\left[ c\left(e^{i\alpha_n^I} - e^{i\alpha_n^{II}} \right) \ket{u_n} - s\left(e^{i\alpha_{n+1}^I} -  e^{i\alpha_{n+1}^{II}} \right) \ket{u_{n+1}} \right].
\end{equation}
 We find that the populations of each spin and each band are 
 \begin{align}    
     p_{n,\uparrow} &= c^2 \cos^2{\frac{\Delta \alpha_n}{2}}, \label{popeqn1}  \\
     p_{n+1,\uparrow} &= s^2 \cos^2{\frac{\Delta \alpha_{n+1}}{2}}, \label{popeqn2}\\
     p_{n,\downarrow} &= c^2 \sin^2{\frac{\Delta \alpha_n}{2}}, \label{popeqn3}\\
     p_{n+1,\downarrow} &= s^2 \sin^2{\frac{\Delta \alpha_{n+1}}{2}}. \label{popeqn4}
 \end{align}

In Fig.~\ref{wavefuncevol} we demonstrate how the wavefunction evolves along each path, both in the general case and for a specific example (in which we do not display dynamic phases for simplicity). In Fig.~\ref{wavefuncevol}b, we find that $\alpha_{n}^{I}=\pi$, $\alpha_{n+1}^{I}=0$, $\alpha_{n}^{II}=\pi$, and $\alpha_{n_1}^{II}=\pi$. Therefore $\Delta \alpha_{n+1} - \Delta \alpha_{n} = \pi-\pi=0$, indicating that along this path joining the nodes, the singularities have opposite frame charges and can annihilate.

In the wavefunctions above, we have assumed that the magnitude of the frame rotation angle $\phi^{\mathrm{frame}}$ is the same at both nodes, $|\phi^{\mathrm{frame}}_I|=|\phi^{\mathrm{frame}}_{II}|$. By studying the winding about a single node using e.g.~the techniques of Ref.~\cite{Brown2022}, the turning angles $\phi^{\mathrm{frame}}$ at each node can be chosen such that this is satisfied, making our interferometry generically applicable even under reduced lattice symmetries.

\subsubsection{Accounting for Dynamic Phases}

Whilst the phases above consist of both a dynamic part $\alpha'$ and a geometric part $\alpha=\alpha^{geo}$, $\Delta \alpha_n^{\mathrm{tot}} = \Delta \alpha_n + \Delta \alpha_n'$, our goal is to extract the geometric part only. When there is sufficient symmetry (for example for the paths shown in Fig.~2 of the main text), the dynamic phase evolved is the same along both paths and don't appear in $\Delta \alpha_n^{\mathrm{tot}}$. More generally, however, we must account for their presence. Once potential method is described here.

By measuring spin populations separately in bands $n$ and $n+1$~\cite{Greiner01_PRL_phasecoherence}, $\cos^2{\left(\Delta \alpha_n^{\mathrm{tot}} /2\right)}$ and $\cos^2{\left(\Delta \alpha_{n+1}^{\mathrm{tot}} /2\right)}$ can be determined. This can also be achieved by measuring spin populations only, if we separately allow an additional $\pi$ phase shift to develop between the two paths (for example by applying a uniform magnetic field) before the final $\pi/2$-pulse. We now seek to determine the geometric contribution, which may be either $0$ or $\pi$, to these phases.

We construct a `reference interferometry loop' technique which relies on repeating the interferometry with twice the acceleration, halving the time for the total process. As a result, the dynamic phase contribution is halved, whilst the geometric phase remains fixed, therefore determining $\cos^2{\left(\Delta \alpha'_{n(+1)}/4 + \Delta \alpha_{n(+1)} /2\right)}$. 

The problem is then reduced to determining whether $\alpha = 0$ or $\pi$ when we have $x=\cos^2\left(\alpha'/2 + \alpha/2 \right)$ and $y=\cos^2\left(\alpha'/4 + \alpha/2 \right)$ (we here drop the $\Delta$ and subscripts for clarity). This can be achieved e.g.~by calculating $x-(2y-1)^2=\sin{\left(\alpha/2\right)}\sin{\left(\alpha'+3\alpha/2\right)}$. Therefore when $\alpha=0$, $x-(2y-1)^2=0$, and generally this quantity will be non-zero when $\alpha=\pi$. We note that when $\alpha'=\pm \pi/2$, this will yield an `accidental zero'. This can be resolved by performing the reference interferometry with four times the acceleration, yielding $z=\cos^2\left(\alpha'/8 + \alpha/2 \right)$, and $y-(2z-1)^2=\sin{\left(\alpha/2\right)}\sin{\left(\alpha'/2+3\alpha/2\right)}$ can be calculated. If this is also $0$, then one can be certain that the geometric phase $\alpha=0$; if we previously had an `accidental zero', then we expect $1/\sqrt{2}$. Indeed, this set of measurements provides a reasonably robust answer. This is because $\mathrm{min}(\mathrm{abs}(x-(2y-1)^2) + \mathrm{abs}(y-(2z-1)^2))=1/\sqrt{2}$ when $\alpha=\pi$, providing a significant `gap' to the value expected for $\alpha=0$ (i.e. $0$).

In this manner, the geometric contributions $\Delta \alpha_n$ and $\Delta \alpha_{n+1}$ can be determined, and used to find the relative charge of the nodes.
%%%%%%%%%%%%%%%%%%%%%%

\subsection{IV. Details for Method 2: Consecutive Deflection}
\label{m2details}

Suppose that we start in a state purely in band $\ket{u_n}$. When we evolve this in and out of a node at an angle such that the frame has rotated by $\phi^{\mathrm{frame}}_1$, upon exiting the node the state is given by
\begin{equation}
    \cos{\phi^{\mathrm{frame}}_1} \ket{u_n} - \sin{\phi^{\mathrm{frame}}_1} \ket{u_{n+1}}.
\end{equation}
If we now evolve this state adiabatically towards the second node, a relative dynamic phase $\alpha'$ will appear between the bands, such that the state (up to some overall phase which can be omitted) just before entering the second node is
\begin{equation}
    e^{i\alpha'} \cos{\phi^{\mathrm{frame}}_1} \ket{u_n} - \sin{\phi^{\mathrm{frame}}_1} \ket{u_{n+1}}.
\end{equation}
Upon entering and exiting this second node with a frame rotation angle $\phi^{\mathrm{frame}}_2$, we have (denoting $\cos{\phi^{\mathrm{frame}}_i}=c_i, \sin{\phi^{\mathrm{frame}}_i}=s_i$)
\begin{equation}
    \left[e^{i\alpha} c_1 c_2 - s_1 s_2\right] \ket{u_n} - \left[e^{i\alpha} c_1 s_2 + s_1 c_2\right] \ket{u_{n+1}}.
\end{equation}
The final population in band $n$ is therefore 
\begin{equation}
    n_1=|e^{i\alpha} c_1 c_2 - s_1 s_2 |^2 = \cos^2{\left(\phi^{\mathrm{frame}}_1+\phi^{\mathrm{frame}}_2\right)} + \sin{2\phi^{\mathrm{frame}}_1} \sin{2\phi^{\mathrm{frame}}_2} \sin^2{\left(\alpha/2\right)}.
\end{equation}
With some algebra we can rewrite this in a more transparent form
\begin{align}
    n_1 &= \frac{1}{2} \left[\cos{\left(2\phi^{\mathrm{frame}}_1 + 2\phi^{\mathrm{frame}}_2 \right)} + 1 + 2\sin{2\phi^{\mathrm{frame}}_1}\sin{2\phi^{\mathrm{frame}}_2}\sin^2{\left(\alpha/2\right)}\right] \\
      &= \frac{1}{2}\left[\cos{2\phi^{\mathrm{frame}}_1} \cos{2\phi^{\mathrm{frame}}_2} - \cos{\alpha} \sin{2\phi^{\mathrm{frame}}_1}\sin{2\phi^{\mathrm{frame}}_2} +1 \right] \\
      &=\frac{1}{2} \left[A \cos{\left(2\phi^{\mathrm{frame}}_1 - \beta\right)} +1 \right]
\end{align}
where
\begin{align}   
    A^2 = \cos^2{\left(2\phi^{\mathrm{frame}}_2\right)}+\cos^2{\left(\alpha\right)}\sin^2{\left(2\phi^{\mathrm{frame}}_2\right)},\label{amplitude} \\
    \tan{\beta}=-\cos{\left(\alpha\right)} \tan(2\phi^{\mathrm{frame}}_2)
\end{align}
as given in the main text.

\subsection{V. Numerical calculations}
\label{numerics}
% Here we describe the numerical validation for the time-evolution along paths in the BZ.

% When an external force $\bm{F}$ is applied to particle which otherwise has Hamiltonian $H$, the Schrödinger equation is
% \begin{equation}
% \label{schrodinger}
%     i \partial_t \ket{\psi} = (H- \bm{F} \cdot \bm{r}) \ket{\psi}.
% \end{equation}
% We now seek to write this in the basis of Bloch wavefunctions. Suppose the initial state has definite (quasi) momentum $\bk$
% \begin{equation}
%     \ket{\psi}=e^{i \bk\cdot \bm{r}} \ket{u_{k}},
% \end{equation}
% where $\ket{u_k}$ is a superposition of bands
% \begin{equation}
%     \ket{u_k}=\sum_n c_n \ket{u_{n,k}}.
% \end{equation}
% The effect of the force is that $\dot{\bk}=\bm{F}$ \cite{Chong2010}, such that through an applied force we can choose the path through momentum-space $\bk(t)$. Therefore, the wavefunction will evolve as
% \begin{equation}
%     \ket{\psi(t)}= e^{i \bk(t) \cdot \bm{r}} \ket{u_{k(t)}},
% \end{equation}
% with 
% \begin{equation}
%     \ket{u_{k(t)}} = \sum_n c_n (t) \ket{u_{n,k(t)}}.
% \end{equation}

% Substituting this into Eqn.~(\ref{schrodinger}), we obtain a form of Schrödinger's equation which acts within the Hilbert space of Bloch states
% \begin{equation}
% \label{schrodingerbloch}
%     i \partial_t \ket{u (t)} = H_{k(t)} \ket{u(t)},
% \end{equation}
% where 
% \begin{equation}
%     H_k = e^{-i \bk \cdot \bm{r}} H e^{i \bk \cdot \bm{r}}
% \end{equation}
% is the Bloch Hamiltonian, and $\bk(t)$ now simply acts as a varying external parameter.

In this section we present numerical calculations to verify that the phases and band populations are as expected from the theory, in addition to further explanations and examples to further clarify the physics. 

We define the path of interest $\bk(t)$, and solve Eqn.~\ref{schrodingerbloch} numerically using Runge-Kutta \cite{Scipy}, for an initial state being a pure band eigenstate at $\bk(0)$. To confirm the results of the interferometry, we must find the phase difference along two paths for a given band. We did this in two different ways. First, if the final states along paths $I$ and $II$ are given by $\ket{u_{I}}$ and $\ket{u_{II}}$, and we would like to find the phase difference of the component in band $\ket{u_n}$, we evaluate the argument of $\braket{u_{I}|u_n}\braket{u_n|u_{II}}$. For example, $\Delta \alpha_{n+1}-\Delta \alpha_n = \braket{u_{I}|u_n}\braket{u_n|u_{II}} - \braket{u_{I}|u_{n+1}}\braket{u_{n+1}|u_{II}}$. This then directly yields the phase differences, which in all cases are as expected. Alternatively, we can follow more closely the experimental procedure. After finding the final states, we construct the wavefunctions for each spin after the $\pi/2$-pulse, and find the corresponding populations. From these, we follow the techniques outlined in Sections~III~and~IV to extract the phase information. 

We confirmed that the techniques are robust to experimental errors in the path and accelerations. For method 1, using the technique described in Section~II, the gap between expected results for similar and opposite frame charges mean that these can still be reliably distinguished. We generically find that the requirement for adiabaticity away from the nodes provides the strongest constraint, but depends on the precise Hamiltonian. For example, for the Kagome lattice we require $\beta \lesssim 1$. Distinguishing relative charges is consistently reliable for randomly distributed $\delta$ with standard deviation $0.05$ (in units with lattice constant $a=1$), where we have also included random $5\%$ error in accuracy of accelerations. On the other hand, when studying the `split gamma' nodes, we require $\beta \lesssim 0.2$, which accordingly forces smaller errors in $\delta$ of order $0.01$, and due to the longer times the accuracy in acceleration must be within $1\%$. Similar considerations apply to method 2; the primary concern is keeping the dynamic phase evolved between the nodes consistent within each run of the experiment, which also requires maintaining adiabaticity away from the nodes.

The above requirements can be compared to parameters achieved in previous experiments \cite{Brown2022}. A difficulty is that decoherence occurs on a timescale of order $\sim 1~\mathrm{ms}$, which puts a lower limit on the acceleration $\beta \gtrsim 1$. This indicates that the techniques developed here are right on the limit of what can be achieved with state of the art experiments. The conditions are easiest when dynamic phases cancel along the two paths, since reference interferometry loops need not be performed, making the Kagome lattice an ideal setting for initial experiments. It should also be noted that reference loops can be avoided if the energies along the path are known.

% \note{Also comment on can be easier with knowing energies. Lastly, just so you know how I've come up with these - I essentially just used that the energies they say are on the order of meV}

\subsection{V A. Results and Examples for Method 1}
\label{m1app}

% The populations were calculated from these using Eqns.~\ref{popeqn1}-\ref{popeqn4}. 
% We worked backwards from these populations, using the techniques to deal with dynamic phases that were described in both the main text (for Method 2) and supplementary materials (for Method 1) to recover the geometric phase contributions. For both methods, to achieve adiabaticity we assumed timescales of $t\gtrapprox 10^2 J^{-1}$, which typically corresponded to a total phase evolution $\mathcal{O}(10^2)$.

\subsubsection{Static Kagome with NN and NNN tunneling terms}

First consider the paths shown in Fig.~2, for the nearest-neighbour Kagome model introduced in the main text. We numerically find phase differences between the paths in band 1 and band 2 as shown in Table~\ref{table:kagnumericalresults}, which are as expected analytically.

\begin{table}[h]
\begin{tabular}{ |c|c|c| } 
\hline
 Path & $\Delta \alpha_1$  & $\Delta \alpha_2$\\
  \hline
 (a1) & $0$ & $0$ \\
 \hline
 (a2) & $0$ & $0$ \\
 \hline
 (b1) & $0$ & $\pi$ \\
 \hline
 (b1) & $\pi$ & $0$ \\
 \hline
\end{tabular}
\caption{Geometric phase differences between the paths for the components in the starting band 1, $\Delta \alpha_1$, and band 2, $\Delta \alpha_2$, for the static nearest-neighbour Kagome model along the paths shown in Fig.~2. Here, due to the $\mathcal{C}_6$ symmetry there is no difference in dynamic phases. All phases have been confirmed numerically, and are in-line with analytic predictions. }
\label{table:kagnumericalresults}
\end{table}

\begin{figure}[h]
\includegraphics[width=0.85\linewidth]{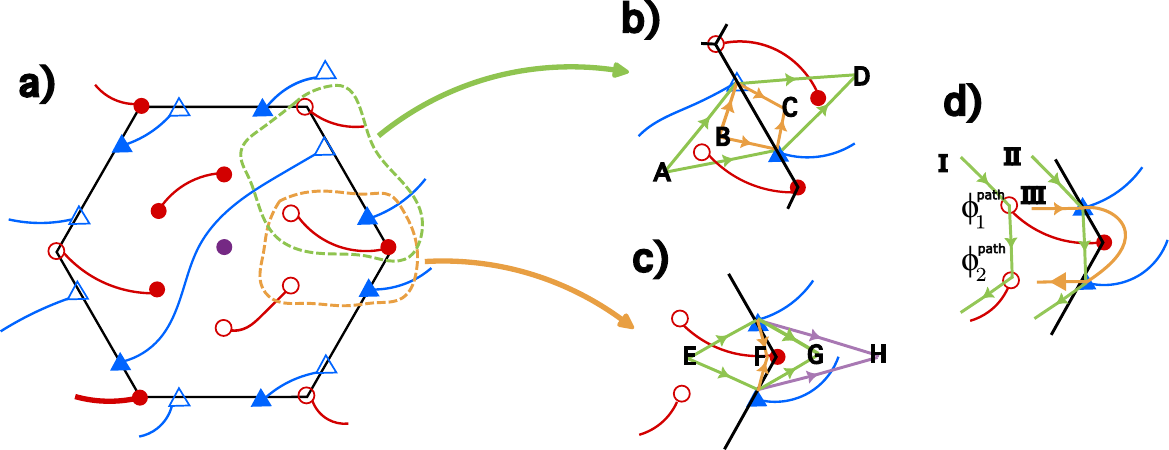}
\caption{a) Location of nodes in the Brillouin zone for the Kagome lattice with next-nearest-neighbour hopping. Blue (triangular) nodes are between bands 1 and 2, whilst red (circular) nodes are between bands 2 and 3. The central $\Gamma$ node is also between bands 2 and 3. Here, the next-nearest neighbour hopping strength $t'$ is related to the nearest-neighbour hopping $J$ through $t'=0.71875 J$. b) and c) Possible paths for interferometry of the nodes between bands 1 and 2. d) Possible paths for a consecutive deflection approach to determining relative charge.}  
\label{nnnkagpaths}
\end{figure}

We further studied interferometry of nodes for a Kagome lattice including next-nearest-neighbour (NNN) hopping terms of amplitude $t'=0.71875 J$ (see supplementary information of Ref.~\cite{Jiang2021acoustic}). In Fig.~\ref{nnnkagpaths}, we demonstrate the location of the nodes in the Brillouin zone for these parameters, with many linear nodes in both gaps. We perform interferometry along paths between the points labelled in Fig.~\ref{nnnkagpaths}, and tabulate the results (confirmed numerically) in Table~\ref{table:nnnkagnumericalresults}. The paths were also numerically verified in the reverse direction, but results are not included in the table. Note that, similar to the nearest-neighbour-only case, the $\mathcal{C}_6$-symmetry means that dynamic phases along each of the chosen paths cancel.

\begin{table}[h]
\begin{tabular}{ |c|c|c| } 
\hline
 Path & $\Delta \alpha_1$  & $\Delta \alpha_2$\\
 \hline
 $\mathrm{A}\to\mathrm{C}$ & $0$ & $0$ \\
 \hline
 $\mathrm{A}\to\mathrm{D}$ & $0$ & $\pi$ \\
 \hline
 $\mathrm{B}\to\mathrm{C}$ & $0$ & $0$ \\
 \hline
 $\mathrm{B}\to\mathrm{D}$ & $0$ & $\pi$ \\
 \hline
 $\mathrm{E}\to\mathrm{F}$ & $0$ & $0$ \\
 \hline
 $\mathrm{E}\to\mathrm{G}$ & $0$ & $\pi$ \\
 \hline
 $\mathrm{E}\to\mathrm{H}$ & $\pi$ & $0$ \\
 \hline
\end{tabular}
\caption{Geometric phase differences between the paths for the components in the starting band 1, $\Delta \alpha_1$, and band 2, $\Delta \alpha_2$, for the Kagome model including next-nearest-neighbour hopping between the points shown in Fig.~\ref{nnnkagpaths}. These results, alongside those for the reversed paths, have been numerically confirmed.}
\label{table:nnnkagnumericalresults}
\end{table}

It is worth lingering on these results briefly, since they demonstrate important points. First (for points shown in Fig.~\ref{nnnkagpaths}), notice that starting at A compared to B makes no difference - here, this is because the Dirac string is between bands 2 and 3 whereas we only have a component of the wavefunction in band 1 at this point. However, had we started in band 2, the Dirac string would have led to a $\pi$ phase shift along one of the paths, changing $\Delta \alpha_1$ and $\Delta \alpha_2$ by $\pi$ (since this phase shift would carry through to the component excited in band 1). This wouldn't change the value of $\Delta \alpha_2 - \Delta \alpha_1$, as we would expect by the arguments in the main text. Physically, this is because the technique measures the patch Euler class along the legs of the interferometry after the node.

Secondly, compare the effect of ending at C to ending at D. When ending at D, a Dirac string between bands 2 and 3 is crossed, leading to a $\pi$ phase shift in band 2. This corresponds to the fact that the patch Euler class of the nodes is $0$ in a region enclosing the path to C, but has magnitude $1$ in a region enclosing the path to D. This is a directly equivalent to the non-Abelian braiding of a node in the gap $1-2$ around a node in gap $2-3$, and physically means that the nodes would annihilate if brought together at C, but would not annihilate if brought together at D. These arguments also apply to ending the path at points F compared to G.

Finally, compare ending at G to ending at H. When we end at H a Dirac string between bands 1 and 2 is crossed, giving a $\pi$ phase shift to the components in both bands. Importantly, this does not affect $\Delta \alpha_2 - \Delta \alpha_1$, which physically corresponds to the fact that braiding around a node in the same gap has no effect.

\subsubsection{Anomalous Floquet Phases}

As a final numerical check of the interferometry technique, we look at anomalous Floquet phases recently introduced in Ref.~\cite{Slager2024}. While out-of-equilibrium settings in general entail intriguing topological phenomena such as topologically protected linking structures~\cite{Tarnowski19_NatCom,Wangchern_17_PRL,zhao2022observation,Unal19_PRR}, periodically driven systems have been attracting a lot of attention in recent years with the anomalous Floquet topological phases that does not exist at equilibrium~\cite{Roy17_PRB,Rudner13_PRX,Unal19_PRL,Wintersperger20_NatPhys}. With multi-gap topological considerations, periodic driving has been recently shown as a versatile tool to move the band singularities within the BZ and braid them~\cite{Slager2024}. Most interestingly, it has been also predicted that by braiding with band nodes appearing in the anomalous gap at the edge Floquet BZ (FBZ) one can obtain an anomalous Euler phase and a anomalous Dirac string phase where {\it all gaps} in the Floquet spectrum including the anomalous one feature either obstructed Euler nodes or Dirac strings~\cite{Slager2024}. These novel phases therefore can be obtained solely under periodic driving and by braiding band singularities in this out-of-equilibrium setting.

\begin{figure}[h]
\includegraphics[width=.6\linewidth]{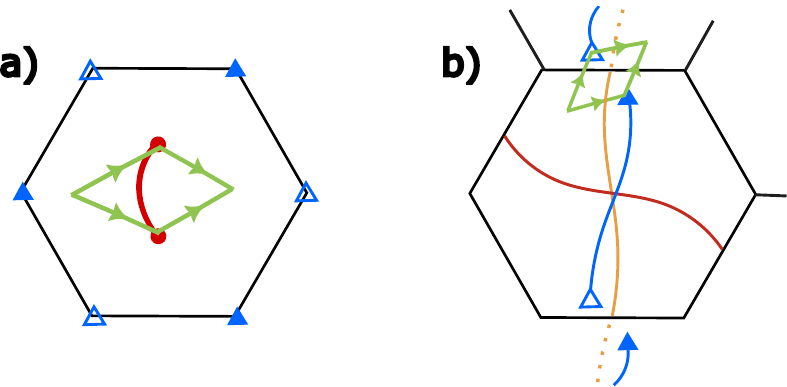}
\caption{Interferometry paths in momentum space for phases induced by Floquet driving. (a) is the `Split Gamma' phase whereby the driving has split the $\Gamma$ node into two linear nodes. (b) is the anomalous Euler phase, where braiding means that the K-nodes now have non-zero patch Euler class.}
\label{floquetpaths}
\end{figure}

We here first consider a weak linear periodic driving that breaks the $\mathcal{C}_6$ symmetry of the Kagome lattice and splits the $\Gamma$ node into two linear nodes, which we dub (i) the `Split Gamma' phase. Secondly, we consider (ii) the `Anomalous Euler' phase, whereby intricate braiding of nodes  through the anomalous Floquet gap results in (K,K') nodes carrying a patch Euler class of magnitude $1$~\cite{Slager2024}. The DSs in the BZ are shown in Fig.~\ref{floquetpaths}, including interferometry paths to study relative charges. Importantly, the fact that these phases no longer possess $\mathcal{C}_6$ symmetry means that the dynamic phases do not cancel in the the chosen paths, and we must take account of these in studying the results. 

To deal with the dynamic phases numerically, we employ two approaches. First, we calculate directly the dynamic phase by integrating the band energies along the paths and subtract it at the end to leave only the geometric contribution. We also work directly from the calculated populations, using the reference interferometry technique described in Sec.~III %~\ref{m1details} 
``Accounting for Dynamic Phases'',
to extract the geometric phase difference. Both methods are numerically confirmed to allow for determining the correct geometric phase difference.

For the `Split Gamma' phase, if we start in band 2 and follow the path drawn in Fig.~\ref{floquetpaths}(a), we obtain $\Delta \alpha_2=0$ and $\Delta \alpha_3=\pi$, indicating a patch Euler class of magnitude 1, as expected. For the `Anomalous Euler' phase, starting in band 1 and following the path in Fig.~\ref{floquetpaths}(b), we obtain $\Delta \alpha_1=0$ and $\Delta \alpha_2=\pi$. This latter result can be inferred from Fig.~\ref{floquetpaths}(ii) in two possible ways. In the gauge that has been chosen for the diagram, the nodes have opposite charge. However, if we pull the Dirac string between bands 2 and 3 through the node on (say) the left, the left node will change sign. In this new gauge it is transparent that the nodes have opposite sign, and the phase differences follow directly. On the other hand, thinking about the interferometry in the gauge in the diagram is also illuminating. The lower path crosses the Dirac string before reaching the node, and since we are in band 1 this therefore has no effect (similarly to the next-nearest-neighbour Kagome, we can also reason that if we had started in band 2, this would like to a $\pi$ phase shift in both components).  The upper path, however, crosses the Dirac string after the node, and so the component in band 2 experiences a $\pi$ phase shift here. We therefore have $\Delta \alpha_2 - \Delta \alpha_1 = \pi$, indicating a patch Euler class of magnitude 1.

\subsection{V B: Results and Examples for Method 2}
\label{m2app}
In this section we present corroborating numerical results for Method 2. In particular, we take the opportunity to step-by-step demonstrate how our techniques work when the winding is not uniform, and emphasise the distinction between the frame-rotation angle $\phi^{\mathrm{frame}}$ and the path rotation angle $\phi^{\mathrm{path}}$.

We have confirmed the technique for nearest-neighbour Kagome in Fig.~3 for similarly- and opposite-charged nodes (K and K'). We first note that the value of $\cos{\alpha'}$ extracted from the `double acceleration' plots (Fig.~3c)  matches up with dynamic phases found by direct integration of energies along the paths, as well as the shifts in the oscillations (Fig.~3b) are quantitatively as expected. In this case, the frames wind uniformly ($\phi^{\mathrm{frame}}=\pm\phi^{\mathrm{frame}}$) and thus choosing $\phi^{\mathrm{frame}}=\pi/2$ fixes $|\phi^{\mathrm{frame}}|=\pi/4$ as required. 

We subsequently confirmed the consecutive deflection technique with the next-nearest-neighbour Kagome model, shown in Fig.~\ref{nnnkagpaths}. The nodes between bands 2 and 3 wind uniformly, and thus are simple to investigate (e.g.~along paths such as Fig.~\ref{nnnkagpaths}(d)(I)). However, the nodes between bands 1 and 2 do not wind uniformly and therefore more care must be taken. We now demonstrate step by step how this can be accounted for in studying path Fig.~\ref{nnnkagpaths}(d)(II).

\begin{figure}[h]
\includegraphics[width=0.5\linewidth]{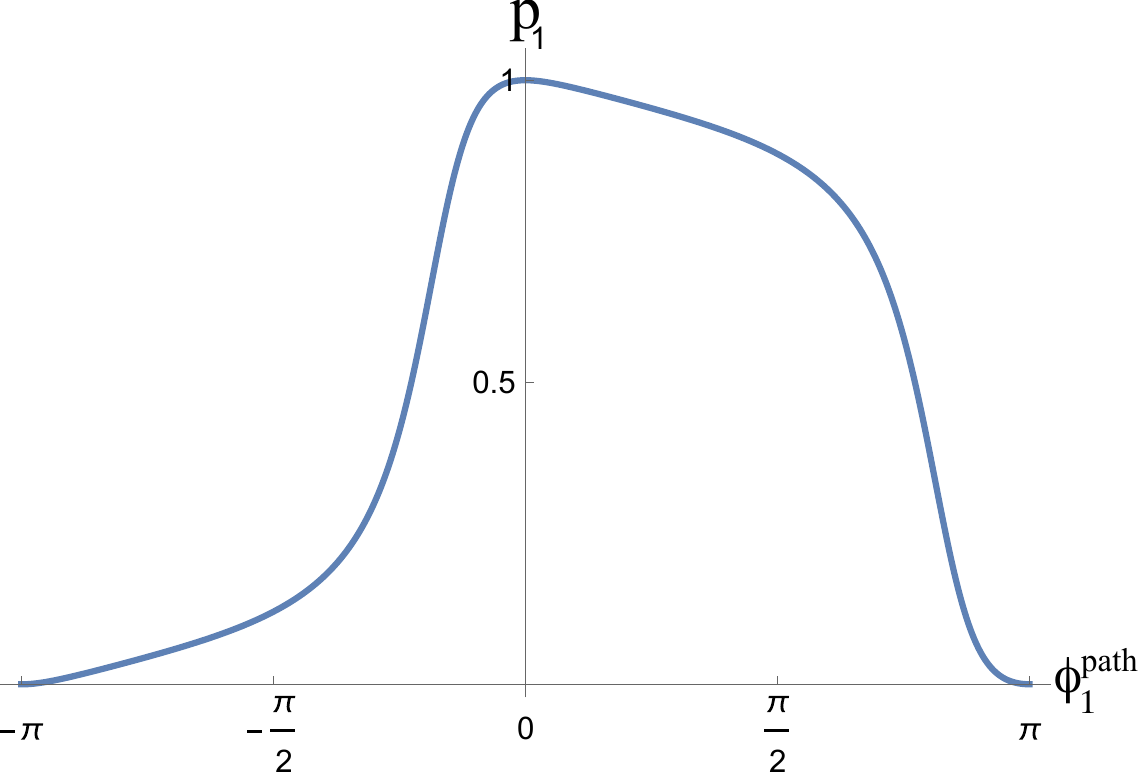}
\caption{Example of a non-uninorm frame winding around a singularity. The population in band 1 after entering and exiting the node at an angle $\phi^{\mathrm{path}}_1$ relative to the path joining the nodes, which corresponds to the frame rotation angle $\phi^{\mathrm{frame}}_1$ through $p_1=\cos^2\phi^{\mathrm{frame}}_1$. This allows the value of $\phi^{\mathrm{path}}_1$ to be chosen such that the frame rotation angle $|\phi^{\mathrm{frame}}|=\pi/4$. Notice that the winding is not uniform, but the net winding is $\pi$ as it is a topologically protected value.}
\label{nnnwinding}
\end{figure}

Step 1: We first must determine the path angles $\phi^{\mathrm{frame}}$ that lead to frame rotation angles $|\phi^{\mathrm{frame}}| = \pi/4$. To do this, we use the methods of Ref.~\cite{Brown2022}. Taking the path which joins the nodes as the fixed point, study the population changes upon entering and exiting each node at an angle. This yields the winding shown in Fig.~\ref{nnnwinding}. From this data, we can read off the values of $\phi^{\mathrm{frame}}$ for which the population is $0.5$, corresponding to $|\phi^{\mathrm{frame}}| = \pi/4$. In this case, we obtain $\phi^{\mathrm{frame}}_1=2.43$ or $\phi^{\mathrm{frame}}_1=-0.71$. Repeating this for the other node, we have $\phi^{\mathrm{frame}}_2=2.43$ (we do not need the negative option here). 

Step 2: For both these values of $\phi^{\mathrm{frame}}_1$, and the fixed $\phi^{\mathrm{frame}}_2$, we prepare a wavepacket in band 1 and accelerate it along the required path with an acceleration $2a$ (where $a$ must be chosen such that the evolution is adiabatic), before measuring the final population in band 1. For the two values of $\phi^{\mathrm{frame}}_1$ we obtain $p_1=0.214$ and $p_1=0.786$. The square of the difference of these values gives us the dynamic phase $\cos^2{\alpha/2}=0.326$. 

Step 3: Finally, we set $\phi^{\mathrm{frame}}_1=2.43$ and $\phi^{\mathrm{frame}}_2=2.43$ (i.e. the $\phi^{\mathrm{frame}}$'s have the same sign), and accelerate the wavepacket along the path with an acceleration $a$. We find that the final population in band 1 is $0.326$, which is the same as $\cos^2{\alpha/2}$, and therefore we know that along this line joining the nodes, the nodes are oppositely charged, and the patch Euler class along this line is 0.

If we instead accelerate the wavepacket along path Fig.~\ref{nnnkagpaths}(d)(III), then we find that the nodes have the same charge. This directly demonstrates the non-Abelian nature of the nodes.

\end{document}